\documentclass{WileyMSP-template}
\usepackage{graphicx}
\usepackage{amssymb}
\usepackage{amsmath, bm}
\usepackage{mathtools, cuted}
\usepackage{color}
\usepackage{indentfirst}
\usepackage{tikz}
\usepackage[labelfont=bf]{caption}

\usepackage{textcomp, gensymb}
\usepackage{array}
\usepackage{booktabs}
\usepackage{threeparttable}
\usepackage{multirow}
\usepackage{makecell}
\usepackage{arydshln}
\usepackage{url} 
\usepackage{ragged2e}
\justifying

\begin{document}

\pagestyle{fancy}
\rhead{\vspace{5mm}}

\title{Ensemble-Embedding Graph Neural Network for Direct Prediction of Optical Spectra from Crystal Structures}

\maketitle


\author{Nguyen Tuan Hung*}
\author{Ryotaro Okabe}
\author{Abhijatmedhi Chotrattanapituk}
\author{Mingda Li*}



\begin{affiliations}
Prof. N. T. H.\\
Frontier Research Institute for Interdisciplinary Sciences, Tohoku University, Sendai 980-8578, Japan\\
Quantum Measurement Group, MIT, Cambridge, MA 02139-4307, USA\\
Email Address: nguyen.tuan.hung.e4@tohoku.ac.jp

R. O.\\
Department of Chemistry, MIT, Cambridge, MA 02139-4307, USA\\
Quantum Measurement Group, MIT, Cambridge, MA 02139-4307, USA

A. C.\\
Department of Electrical Engineering and Computer Science, MIT, Cambridge, MA 02139-4307, USA\\
Quantum Measurement Group, MIT, Cambridge, MA 02139-4307, USA

Prof. M. L.\\
Department of Nuclear Science and Engineering, MIT, Cambridge, MA 02139-4307, USA\\
Quantum Measurement Group, MIT, Cambridge, MA 02139-4307, USA\\
Email Address: mingda@mit.edu

\end{affiliations}



\begin{abstract}

Optical properties in solids, such as refractive index and absorption, hold vast applications ranging from solar panels to sensors, photodetectors, and transparent displays. However, first-principles computation of optical properties from crystal structures is a complex task due to the high convergence criteria and computational cost. Recent progress in machine learning shows promise in predicting material properties, yet predicting optical properties from crystal structures remains challenging due to the lack of efficient atomic embeddings. Here, we introduce GNNOpt, an equivariance graph-neural-network architecture featuring automatic embedding optimization. This enables high-quality optical predictions with a dataset of only 944 materials. GNNOpt predicts all optical properties based on the Kramers-Kr{\"o}nig relations, including absorption coefficient, complex dielectric function, complex refractive index, and reflectance. We apply the trained model to screen photovoltaic materials based on spectroscopic limited maximum efficiency and search for quantum materials based on quantum weight. First-principles calculations validate the efficacy of the GNNOpt model, demonstrating excellent agreement in predicting the optical spectra of unseen materials. The discovery of new quantum materials with high predicted quantum weight, such as SiOs which hosts exotic quasiparticles, demonstrates the potential of GNNOpt in predicting optical properties across a broad range of materials and applications.

\end{abstract}


\section{Introduction}
Understanding the optical properties of materials is crucial for designing and optimizing optoelectronic devices such as LEDs, solar cells, photodetectors, and photonic integrated circuits (PICs). These devices play an indispensable role in the current resurgence of the semiconductor industry~\cite{marpaung2019integrated,bogaerts2020programmable,meng2021optical,priolo2014silicon}. The linear optical responses, in particular, offer insights into fundamental parameters such as energy bandgaps, transparency, reflectivity, and refractive index, which are essential for controlling the light-matter interactions~\cite{liu2023interference,khurgin2022expanding}. Tremendous scientific and industrial interests have driven both experimental and computational efforts toward high-throughput screening of candidate materials for tailored optical applications. Current experimental techniques, including ellipsometry, UV-Vis spectroscopy, and Fourier transform infrared spectroscopy (FTIS), are commonly used to obtain materials' optical spectra. However, they are limited to specific wavelength ranges and often require stringent sample conditions, making them unideal for high-throughput material screening
~\cite{gulo2023exploring,saito2024deep,smith2011fundamentals,perkampus2013uv}. On the other hand, first-principles calculations based on density functional theory (DFT) can compute optical spectra across all wavelength ranges~\cite{hung2022quantum}. However, DFT requires dense k-point sampling for the convergence of complex dielectric functions or absorption coefficients. For instance, in graphite, the transition of excited electrons by light occurs near the Dirac point, necessitating a k-point sampling of over $100,000$ k-points to accurately capture the optical transitions, resulting in a time-consuming calculation~\cite{gulo2022exploring}. Consequently, high-throughput DFT calculations for optical spectra are largely limited to materials with a small number of atoms per unit cell~\cite{yang2022high}.

Machine-learning methods are increasingly being adopted in materials research to accelerate materials discovery through high-throughput property prediction~\cite{liu2017materials,genty2021machine,pollice2021data}. One successful approach is the use of graph neural networks (GNNs) to predict material properties directly from crystal structures~\cite{xie2018crystal,scarselli2008graph,chen2021direct}. 
Chen \textit{et al.}~\cite{chen2021direct} build a GNN model using an equivariant neural network with E(3)NN~\cite{geiger2022e3nn}, which predicts the phonon density-of-states (DOS) by using only atomic species and positions as input parameters. The success of machine learning models in structure-property prediction suggests a potential for applying machine learning to optical spectra prediction. However, challenges remain in developing an effective model that not only accurately predicts optical properties from a small available database, but also extracts useful information from complex relationships in optical spectra, such as the Kramers–Kr{\"o}nig (K-K) relations and the $f$-sum rule~\cite{lucarini2005kramers}. Such a model could be instrumental in searching for photovoltaic materials for energy conversion or in understanding fundamental physics through the optical spectra of materials.

In this work, we develop GNNOpt, a GNN model that establishes a direct relationship between crystal structures and frequency-dependent optical properties. Rather than focusing on building a more complex neural network structure, we emphasize automatic embedding optimization on top of the equivariant neural networks. This approach surpasses the commonly-used fixed embedding schemes, and goes beyond the feature selection and importance method by integrating different features through additional neural networks. This enables high-quality optical predictions with a small dataset of 944 materials~\cite{yang2022high,jain2013commentary}. In a GNN model, converting crystal structure into machine-readable graph representations is essential. While the distance vector between an atom and its neighbor is a mandatory fixed embedding for applying the equivariant neural networks in E(3)NN~\cite{geiger2022e3nn}, particularly as input parameters of the spherical harmonics in a tensor product, the feature embedding, or representation of an atom, serve as initial parameters and often relies on human intuition to select the best descriptors in a GNN. For phonon DOS predictions, atomic mass is a natural embedding~\cite{chen2021direct,okabe2023virtual}. However,  atomic mass is less relevant for optical properties, and identifying suitable atomic embeddings can be challenging since there are various physical, chemical, structural, and environmental descriptors available. To address this challenge, we propose an ensemble embedding layer for the GNN model. The ensemble embedding layer assigns a learnable weight to each feature embedding. As a result, the GNN model can automatically identify the most important descriptors for specific physical properties from several selected descriptors. Combining the ensemble embedding and the equivariant neural networks, our GNN model can directly and accurately predict optical spectra using only crystal structure as input, which shows superior performance than any fixed embeddings. Additionally, thanks to the K-K relation, we were able to extract all frequency-dependent optical spectra, such as the absorption coefficient, the refractive index, and reflectivity, from only the real (or imaginary) part of the dielectric function. Thus, our predictive model can capture the main features of all-optical spectra, even for crystal structures with unseen elements. By predicting the optical spectra in 5,281 unseen crystal structures, we identify a list of high-performing materials for solar cell application, supported by additional first-principles calculations. The $f$-sum rule is another universal constraint in the linear optical response, which determines the quantum characteristics of a material by integrating over an infinite spectral range. A recent theory shows that quantum weight, a parameter based on the variation of $f$-sum rule~\cite{onishi2024fundamental}, is directly connected to the ground state quantum geometry and topology. Using quantum weight, our GNNOpt model successfully identifies several quantum materials with high quantum weight, such as SiOs, validated by first-principles calculations. Thus, our work presents an efficient approach to obtaining optical spectra directly from crystal structures, offering diverse applications in materials science.

\section{GNNOpt: A machine learning model for optical spectra prediction}

\begin{figure}[ht!]
    \centering
    \includegraphics[width=0.95\linewidth]{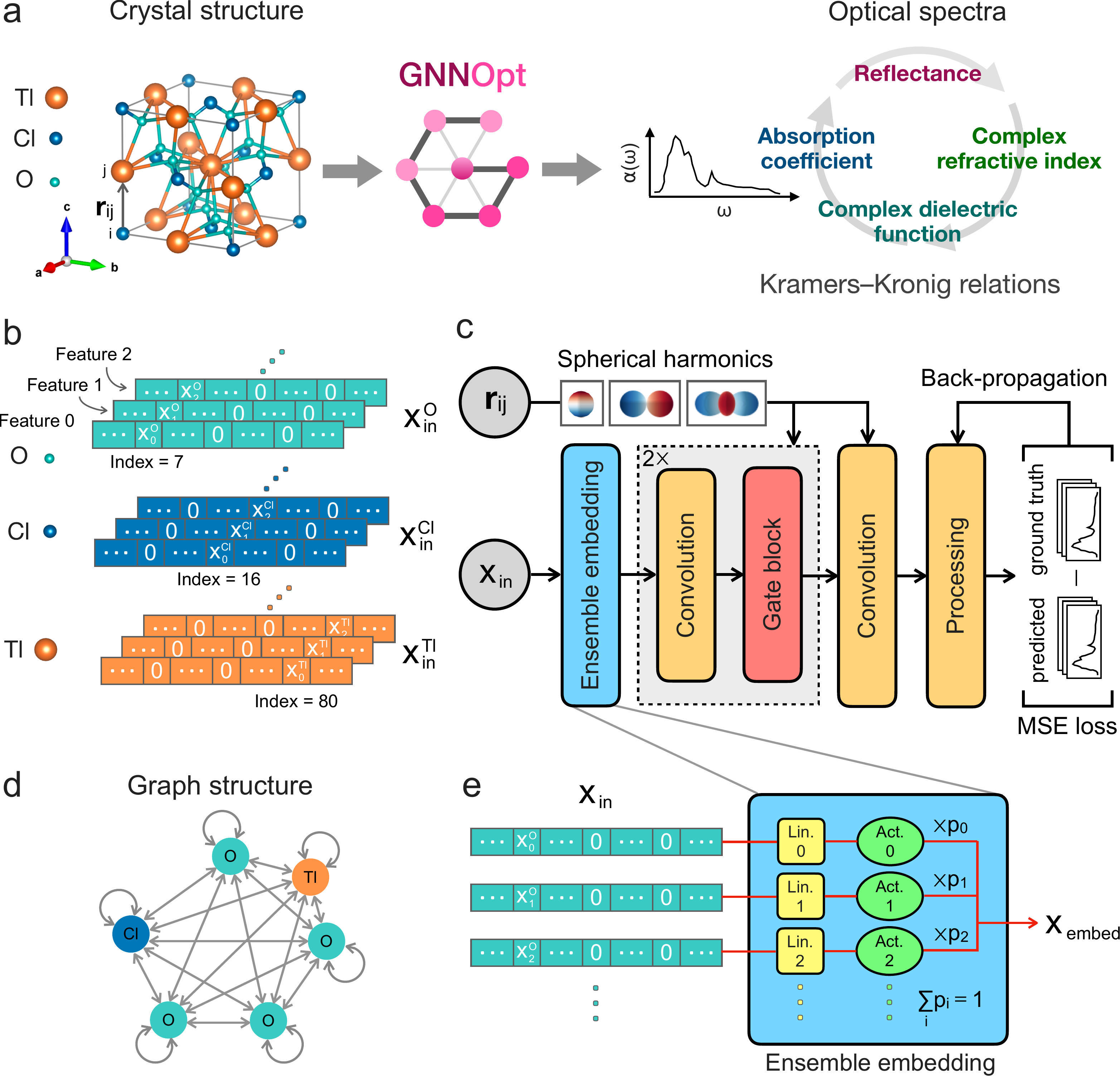}
    \caption{\textbf{The GNNOpt for optical spectra prediction.} \textbf{a.} The crystal structure is the only input parameter needed for the GNNOpt to predict the optical spectra, taking TlClO$_4$ as an example. $\bm{r}_{ij}$ is a distance vector between $i$-th and $j$-th atoms, and a number of frequency-dependent optical properties can be predicted. \textbf{b.}  Input features of each atomic species (O, Cl, and Tl) in feature-weighted one-hot representation, which is a 118-element-long array of zeroes except an element with an index equal to the atomic number where the value is equal to the feature. \textbf{c.}  Overview of the GNNOpt architecture. All atomic input features are automatically optimized by an ensemble embedding layer. Then, the embedded features are passed through a sequence of equivariant graph convolution and gated nonlinear layers parameterized with $\bm{r}_{ij}$'s radial and spherical harmonic representations. After that, the result is passed to a post-process layer, including activation and aggregation, to generate the predicted output spectra. Finally, the network weights are trained by minimizing the mean-square error (MSE) loss function between the predicted and ground-truth spectra. \textbf{d.}  Graph representation of periodic lattice of TlClO$_4$, where the graph nodes represent the atoms inside a unit cell, and edges represent message passing directions of graph convolutions. \textbf{e.} In an ensemble embedding layer, each atomic feature is independently embedded with its linear layer (Lin. 0, Lin. 1, ...) and activation layers (Act. 0, Act. 1, ...). After that, all embedded features are weighted and averaged by learnable weight probability $ p_i$. The same ensemble embedding layer's parameter values are used across all elements, allowing interpretability of the model feature importance.}
    \label{fig:model}
\end{figure}

The input and output of the GNNOpt are shown in  \textbf{Figure~\ref{fig:model}a}, in which the crystal structure is only the input parameter for the GNNOpt and the optical spectra including complex dielectric function ($\epsilon_1(\omega)$ is the real part and $\epsilon_2(\omega)$ is the imaginary part), absorption coefficient ($\alpha(\omega)$), complex refractive index ($n(\omega)$ is the real part and $k(\omega)$ is the imaginary part), and reflectance ($R(\omega)$), where $\omega$ is the angular frequency of light and the photon energy is given by $E=\hbar\omega$, where $\hbar$ is the reduced Planck constant. All optical spectra are related together by the Kramers-Kr{\"o}nig relations, which will be discussed in the next paragraph. In \textbf{Figure~\ref{fig:model}b}, we show the input features of each atomic species (or each node), in which we adopt three features, including the atomic mass ($x_0$), dipole polarizability ($x_1$), and the effective covalent radius ($x_2$). All input features use a one-hot encoding to denote the atomic species. For example, the oxygen atom is encoded as $[0,\ldots,x_0^\text{O},\ldots,0]$ for the feature 0 (i.e., atomic mass) with a 118-element-long array of zeroes except for an element with an index $= 8$, which equals to the atomic number of the oxygen minus one. Similarly, we have $[0,\ldots,x_1^\text{O},\ldots,0]$ and $[0,\ldots,x_2^\text{O},\ldots,0]$ for the feature 1 (dipole polarizability) and the feature 2 (covalent radius), respectively. Thus, the one-hot representation of the oxygen atom with three features is $\bm{x}_{\text{in}}^\text{O}=([0,\ldots,x_0^\text{O},\ldots,0]; [0,\ldots,x_1^\text{O},\ldots,0]; [0,\ldots,x_2^\text{O},\ldots,0])$. After that, all input features of each atom are automatically optimized for their embedding mixture by an ensemble embedding layer, as shown in \textbf{Figure~\ref{fig:model}c}. Then, the embedded features are passed through a sequence of equivariant graph convolution and gated nonlinear layers parameterized with the input parameter $\bm{r}_{ij}$. Here, $\bm{r}_{ij}$ is a distance vector between $i$-th atom and neighbor $j$-th atoms up to a radial cut-oﬀ value $r_\text{cut} = 6$ \AA\ (see \textbf{Figure~\ref{fig:model}a}). Here, the periodic boundary condition is considered when constructing a graph. To achieve the equivariance, the convolutional filters are designed to be composed of learnable radial functions $R(|\bm{r}_{ij}|)$ and spherical harmonics $Y_\ell^m(\bm{r}_{ij}/|\bm{r}_{ij}|)$, where the indices $\ell$ and $m$ indicate the degree and order of the function. Therefore, the geometric information and all crystallographic symmetries of input crystal structures are preserved in GNNOpt. After the final convolution layer, all resulting features are summed and passed through a processing layer, including activation (ReLU) and normalization (using the median value of ground truth) steps to predict the optical spectra. The weights of the GNNOpt are optimized by minimizing the mean squared error (MSE) loss function between the predicted and ground truth spectra. Figure~\ref{fig:model}d shows the graph representation of a unit-cell TlClO$_4$, in which nodes represent atoms of a unit cell, and edges represent the message passing direction of graph convolutions layers. In \textbf{Figure~\ref{fig:model}e}, we show the detail of the ensemble embedding layer, which is the key to performance improvement even without any neural network model change. For each atom, each feature is independently embedded with its linear and activation layers. After that, all embedded features are evaluated by a weighted average by learnable mixing probability $p_i$, in which $p_i$ are normalized by $\sum_i p_i = 1$ (see Method Section for the full hyperparameters of the GNNOpt).

\begin{figure}
    \centering
    \includegraphics[width=0.90\linewidth]{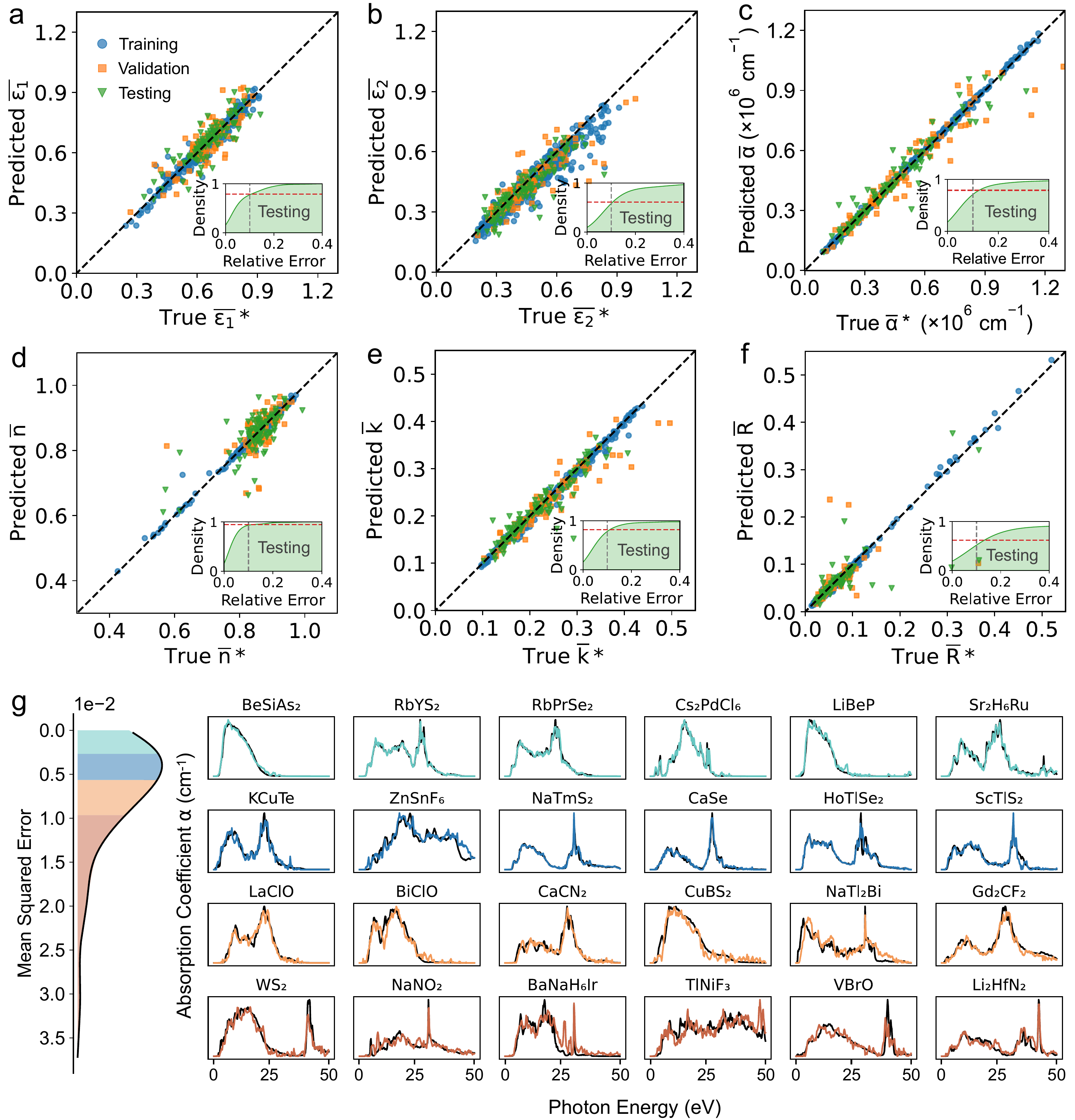}
    \caption{\textbf{Performance of the GNNOpt in predicting different optical properties.} \textbf{a--f.} Weighted average magnitudes of a few optical properties predicted from the GNNOpt model, $\overline{W}$, compared with the ground-truth (true), $\overline{W}^*$, with the $\overline{W}=\overline{\epsilon_1}, \overline{\epsilon_2}, \overline{\alpha}, \overline{n}, \overline{k},$ and $\overline{R}$, for complex dielectric function, absorption coefficient, complex refractive index, and reflectance, respectively. The data points of circle (blue), square (orange), and triangle (green) denote the training, validation, and testing datasets, respectively. The inset shows the cumulative kernel-density-estimator (KDE) plot of the relative error, $|\overline{W}-\overline{W}^*|/\overline{W}^*$, of the only testing dataset. \textbf{g.} Mean squared error (MSE) distribution and 24 randomly selected materials in the test dataset corresponding to each error quartile in the MSE distribution for absorption coefficient $\alpha(\omega)$. Color lines are the GNNOpt-predicted spectra, and black lines are the DFT ground-truth spectra.}
    \label{fig:gnnopt}
\end{figure}

Next, we apply GNNOpt to predict the optical spectra given the limited training data with 944 materials, which are calculated by DFT within the independent-particle approximation (IPA)~\cite{yang2022high}. The database is obtained from the Material Project~\cite{jain2013commentary}. The details for the data preparation and statistics are given in Section 7.1 and \textbf{Figure S1} in the Supporting Information. We note that the linear optical properties are not independent but follow the K-K relations~\cite{lucarini2005kramers,hutchings1992kramers}:
\begin{equation}
    \epsilon_1(\omega) = 1 +\frac{2}{\pi}P\int_0^\infty\frac{\epsilon_2(\omega')\omega'}{\omega'^2-\omega^2}\mathrm{d}(\omega'),
\end{equation}
\begin{equation}
    \epsilon_2(\omega) = -\frac{2\omega}{\pi}P\int_0^\infty\frac{\epsilon_2(\omega')\omega'}{\omega'^2-\omega^2}\mathrm{d}(\omega'),
\end{equation}
\begin{equation}
    \alpha(\omega) = \frac{\sqrt{2}\omega}{c}\sqrt{\sqrt{\epsilon_1^2(\omega)+\epsilon_2^2(\omega)}-\epsilon_1(\omega)},
\end{equation}
\begin{equation}
    n(\omega) = \sqrt{\frac{1}{2}\left(\sqrt{\epsilon_1^2(\omega)+\epsilon_2^2(\omega)}+\epsilon_1(\omega)\right)},
\end{equation}
\begin{equation}
    k(\omega) = \sqrt{\frac{1}{2}\left(\sqrt{\epsilon_1^2(\omega)+\epsilon_2^2(\omega)}-\epsilon_1(\omega)\right)},
\end{equation}
\begin{equation}
    R(\omega) = \frac{(n(\omega)-1)^2+k^2(\omega)}{(n(\omega)+1)^2+k^2(\omega)},
\end{equation}
where $P$ in Equations (1) and (2) denotes the Cauchy principal value. The database, which originally only included $\epsilon_1$, $\epsilon_2$, and $\alpha$, is expanded for $n$, $k$, and $R$ by using Equations (4), (5), and (6), respectively. It is noted that only the averaged optical values, i.e., $\alpha=(\alpha_{xx}+\alpha_{yy}+\alpha_{zz})/3$, are available in the database~\cite{yang2022high}.

To compare the predicted spectra with the DFT ground-truth spectra, we calculate the weighted mean, $\overline{W}$, of the optical spectra, $W(\omega)$, by
\begin{equation}
    \overline{W}=\frac{\int  \mathrm{d}{\omega} W(\omega) \omega}{\int \mathrm{d}{\omega} W(\omega)},
\end{equation}
where $W(\omega)$ denotes the optical spectra, such as $\epsilon_1(\omega), \epsilon_2(\omega), \alpha(\omega), n(\omega), k(\omega),$ or $R(\omega)$. The corrections between the GNNOpt-predicted, $\overline{W}$, and DFT ground-truth spectra, $\overline{W}^*$ are plotted in \textbf{Figure~\ref{fig:gnnopt}a--f}. Here, the coefficients of determination, $R^2$, of the test set are 0.72, 0.74, 0.93, 0.51, 0.86, and 0.55 for $\epsilon_1$, $\epsilon_2$, $\alpha$, $n$, $k$, and $R$, respectively, which show excellent agreement for the cases of $\alpha$ and $k$. On the other hand, the relative error $|\overline{W}-\overline{W}^*|/\overline{W}^*$ below 10\% for the test set (see the inset figures) of $\epsilon_1$, $\epsilon_2$, $\alpha$, $n$, $k$, and $R$ are 79\%, 61\%, 79\%, 95\%, 82\%, and 62\%, respectively, which shows the high performance of our model to predict $\epsilon_1$, $\alpha$, $n$, and $k$. To visualize the model performance, in \textbf{Figure~\ref{fig:gnnopt}g}, we also plot $\alpha(\omega)$ of the 24 randomly selected materials from the test set in each mean-squared-error (MSE) quartile, in which the color lines are the GNNOpt-predicted spectra and the black lines are the DFT ground-truth spectra. In \textbf{Figure~\ref{fig:gnnopt}g}, the 1st to 4th rows in the right figure correspond to the 1st to 4th error quartiles in the left figure by the same color. The overlap between the predicted $\alpha(\omega)$ (color lines) and the DFT calculation (black line) in \textbf{Figure~\ref{fig:gnnopt}g} suggests that our model can accurately predict optical spectra, in which all spectrum peaks are reproduced from the DFT calculation even for the 4th row. In the 3rd and 4th rows, some noise can be found in the predicted $\alpha(\omega)$ of NaTl$_2$Bi or TlNiF$_3$. This is because of the origin of the Dirac delta function in the dielectric function formula, which is often solved by adopting a broadening parameter or increasing the number of k-point samples in the DFT calculation~\cite{hung2022quantum}. For the GNNOpt model, a uniform filter of the SciPy library~\cite{virtanen2020scipy} can be applied instead of the broadening parameter in the DFT to reduce the noise of the spectrum. The MSE loss and the compared full spectra of the training, validation, and testing datasets for $\epsilon_1$, $\epsilon_2$, $\alpha$, $n$, $k$, and $R$ are given in \textbf{Figures S4-S10} in the Supporting Information, respectively.

\begin{table*}[t]
\centering
\begin{tabular}{cccccc}
\toprule
\multicolumn{1}{c}{\multirow{2}{*}{\textbf{Optic}}} & \multicolumn{3}{c}{\textbf{Fixed embedding}} & \multicolumn{2}{c}{\textbf{Ensemble embedding} } \\\cmidrule(lr){2-4} \cmidrule(lr){5-6}
\multicolumn{1}{c}{}  & Atomic mass  &  Dipole polarizability  & Covalent radius  &  Ensemble      & Probability ($p_0, p_1, p_2$)  \\\midrule
$\epsilon_1$ &  0.69   &  0.67  &  0.72  &  0.72  &  0.0021, 0.0017, 0.9962  \\\hdashline
$\epsilon_2$ &  0.62   &  0.63  &  0.58  &  0.74  &  0.0026, 0,0002, 0.9972  \\\hdashline
$\alpha$     &  0.93   &  0.93  &  0.91  &  0.93  &  0.0001, 0.0008, 0.9991  \\\hdashline
$n$          &  0.48   &  0.51  &  0.48  &  0.51  &  0.0011, 0.0013, 0.9976  \\\hdashline
$k$          &  0.85   &  0.86  &  0.85  &  0.86  &  0.0001, 0.0002, 0.9997  \\\hdashline
$R$          &  0.61   &  0.48  &  0.57  &  0.55  &  0.0033, 0.0018, 0.9949  \\\bottomrule     
\end{tabular}
 \caption{The coefficient of determination, $R^2$, of the test set with the fixed embedding and ensemble embedding layer. The feature embedding: atomic mass, dipole polarizability, or covalent radius is used in the fixed embedding, while ensemble embedding considers three feature embeddings.}
 \label{tab:1}
\end{table*}

In \textbf{Table~\ref{tab:1}}, we compare the $R^2$ values of the test set with the ensemble embedding layer and fixed embeddings. In this present study, we use three feature embeddings: atomic mass, dipole polarizability, and covalent radius, since these features are provided for all elements in the periodic table. Nevertheless, the ensemble embedding layer can work with any other added features. Table~\ref{tab:1} shows the performance improvement of the GNNOpt with the ensemble embedding layer for $\epsilon_2$. For other optical spectra, the GNNOpt with the ensemble embedding layer is comparable with the best-trained fixed embedding model. On the other hand, the value of $p_i$ of the ensemble embedding layer can tell us which feature is the main contribution. For the GNNOpt, the mixing probability shows that the covalent radius (feature 2) is almost completely dominant with $p_2 \approx 1$. We note that the covalent radius is a measure of the size of an atom that forms part of one covalent bond~\cite{pyykko2009molecular}. On the other hand, the dipole moment, which is related to the optical properties, is defined as the product of the total amount of charge and the covalent bond length. This relationship might explain why the covalent radius is the most dominant feature in the GNNOpt. 

To evaluate the scalability of the GNNOpt model for unseen materials, we perform the GNNOpt with the test set containing larger atomic site numbers $N_\text{test}$ than the training and validation sets. There are there cases of the test set as $N_\text{test}=8,9$ (case I), $N_\text{test}=7,8,9$ (case II), and $N_\text{test}=6,7,8,9$ (case III). As shown in \textbf{Figures S11a-c}, $R^2$ for the test set are 0.83, 0.84, and 0.86 for the cases I, II, and III, respectively. Moreover, the training set contains only 443 materials with $2 \leq N_\text{train} \leq 5$ in case III. However, the prediction optical spectra show a good overlap with the DFT ground truth for the test set with $6 \leq N_\text{test} \leq 9$ (see Figure S11d). Thus, it suggests that the GNNOpt can accurately predict the optical properties of unseen complex materials with larger systems than the training set. 

\section{The Kramers-Kr{\"o}nig relations}

\begin{figure}[t]
    \centering
    \includegraphics[width=0.60\linewidth]{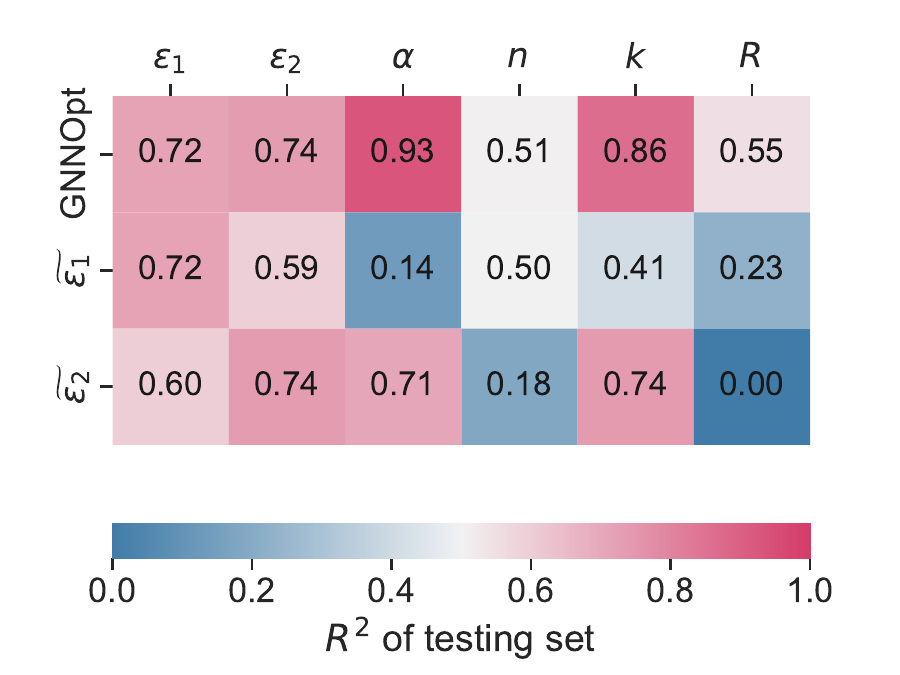}
    \caption{Coefficient of determination $R^2$ of the test set for the optical spectra $W = \epsilon_1, \epsilon_2, \alpha, n, k$, and $R$. For the 1st row, $W$ is directly predicted by using GNNOpt. For the 2nd and 3rd rows, $W$ is calculated from the predicted $\widetilde{\epsilon_1}$ and $\widetilde{\epsilon_2}$ by using the Kramers-Kr{\"o}nig relations, respectively.}
    \label{fig:KK}
\end{figure}

To evaluate the Kramers-Kronig relations for predicted optical properties, first, we use the complex dielectric function $\widetilde{\epsilon_1}$ and $\widetilde{\epsilon_2}$, which are predicted by the GNNOpt, to calculate other optical properties by using Equations (1)-(6). Then, we calculate the $R^2$ coefficients for the test set for each of the optical properties. Finally, we compare these $R^2$ values from the K-K relations with $R^2$ of the corresponding optical properties from GNNOpt, as shown in \textbf{Figure~\ref{fig:KK}}. It notes that we use the same training, validation, and test sets for all cases. The highest $R^2$ value in the 1st row suggests that directly predicting the optical properties is better than calculating the optical properties from the predicted $\widetilde{\epsilon_1}$ (or $\widetilde{\epsilon_2}$). In other words, the K-K relations must be applied before the GNNOpt model to predict optical spectra, i.e., $\epsilon_1 (\epsilon_2) \to$ K-K relations $\to W \to$ GNNOpt $\to \text{predicted-}{W}$.
In the 2nd and 3rd rows of Figure~\ref{fig:KK}, $R^2$ of $\alpha/\widetilde{\epsilon_2}$ ($0.71$) and $k/\widetilde{\epsilon_2}$ ($0.74$) are larger than that of $\alpha/\widetilde{\epsilon_1}$ ($0.14$) and $k/\widetilde{\epsilon_1}$ ($0.41$). This implies that $\alpha$ and $k$ are more dependent on $\epsilon_2$ than $\epsilon_1$, which is clearly explained by Equations (3) and (4), $\alpha(\omega)=2\omega k(\omega)/c \propto \sqrt{|\epsilon_2|}$. In contrast, $R^2$ of $n/\widetilde{\epsilon_1}$ ($0.50$) and $R/\widetilde{\epsilon_1}$ ($0.23$) are larger than that of $n/\widetilde{\epsilon_2}$ ($0.18$) and $R/\widetilde{\epsilon_2}$ ($0.00$). That means that $n$ and $R$ are more dependent on $\epsilon_1$ than $\epsilon_2$. $R^2$ of $\epsilon_1/\widetilde{\epsilon_1}$ (0.72) and $\epsilon_2/\widetilde{\epsilon_2}$ (0.74) are same with that of GNNOpt since $\epsilon_1=\widetilde{\epsilon_1}$ and $\epsilon_2=\widetilde{\epsilon_2}$. By evaluating the K-K relations before and after training the GNNOpt model, we found that the K-K relations should be applied before training the model.

\section{Screening solar cell materials}
The first application of the GNNOpt is to identify potential solar cell materials with high-performance energy conversion. For this task, Shockley-Queisser (SQ) limit is a key factor to estimate the upper limit for solar-energy conversion efficiency~\cite{shockley1961detailed}. However, the SQ limit assumes that the absorption coefficient is a step function or the infinity thickness of the absorber material. An alternative parameter that is more suitable for real PV materials and devices is introduced by Yu and Zunger~\cite{yu2012identification}, which is the spectroscopic limited maximum eﬃciency (SLME). For the SLME method, the efficiency of the energy conversion, $\eta$, is defined as the ratio between the maximum output power density, $P_{\text{out}}$, and the incident solar power density, $P_{\text{solar}}$,
\begin{equation}
    \eta = \frac{P_{\text{max}}}{P_{\text{solar}}}.
\end{equation}
$P_{\text{max}}$ and $P_{\text{solar}}$ can be written in terms of $J-V$ characteristic of the solar cell and the solar spectrum, respectively~\cite{yu2012identification}:
\begin{equation}
    P_{\text{max}}= \text{max}\{JV\}_V = \text{max}\{[J_\text{sc}-J_0(e^{qV/k_BT}-1)]V\}_V,
\end{equation}
and
\begin{equation}
    P_{\text{solar}}= \int_0^\infty E I_\text{solar}(E) \mathrm{d}E,
\end{equation}
respectively, where $J$ is the total current density, $V$ is potential over the absorber layer, $k_B$ is the Boltzmann constant, $T$ is the temperature of the solar device, $q$ is the elementary charge. $J_\text{sc}$ and $J_0$ are the short-circuit current density and the reverse saturation current density, respectively, which depend on the absorption coefficient, $\alpha(E)$, and the thickness, $L$, of the material as follows~\cite{yu2012identification}:
\begin{equation}
    J_\text{sc}=q\int_0^\infty \left[1- e^{-2\alpha(E)L}\right]I_\text{solar}(E)\mathrm{d}E
\end{equation}
and
\begin{equation}
    J_0=\frac{J_0^r}{f_r}=\frac{q\pi}{f_r}\int_0^\infty \left[1- e^{-2\alpha(E)L}\right]I_\text{bb}(E,T)\mathrm{d}E,
\end{equation}
where $I_\text{solar}(E)$ is the AM1.5G solar spectrum, $I_\text{bb}(E,T)$ is the black-body spectrum, and $f_r$ is the radiative recombination current density. From Equations (8)-(12), the material property-related inputs $\alpha(E)$, $L$, $T$, and $f_r$ are required to calculate the SLME. In this work, we assume the $L = 500$ nm, $T=300$ K, and the radiative recombination is the
only recombination process, i.e., $f_r=1$, which is a good approximation for the materials where radiative recombination dominates, such as GaAs~\cite{schnitzer1992ultra}. Since the solar spectrum for solar energy harvesting ranges from $200-2500$ nm (i.e., $0.5-6.0$ eV), we thus retrain the GNNOpt model with the photon energy $< 10$ eV, which is sufficient for the SLME calculation.

\begin{figure}[t]
    \centering
    \includegraphics[width=0.90\linewidth]{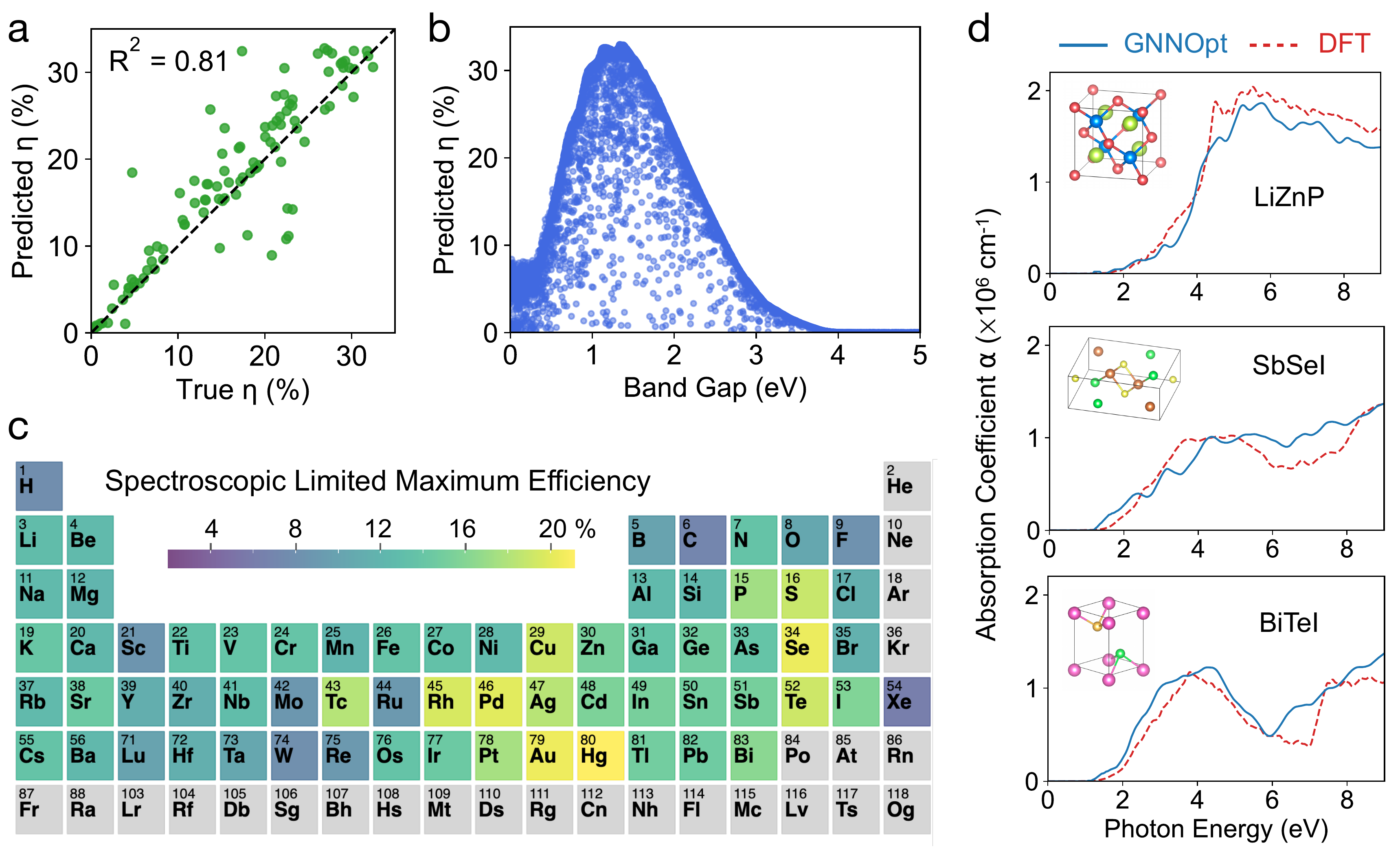}
    \caption{\textbf{Searching material candidates for solar energy harvesting applications with GNNOpt.} \textbf{a.} Comparison between predicted and true efficiencies $\eta$, obtained by the GNNOpt and the DFT ground truth, respectively, for the test set. Here, $\eta$ is estimated by the spectroscopic limited maximum efficiency (SLME). \textbf{b.} $\eta$-predicted by the GNNOpt of the 5,281 stable materials from the Material Project, which is plotted as a function of the energy band gap. \textbf{c.} The periodic table is colored by the SLME of materials containing each element. \textbf{d.} Absorption coefficients $\alpha(E)$ are plotted as a function of photon energy $E$ by using the GNNOpt (solid lines) and the DFT (dashed lines) for the unseen materials, including LiZnP, SbSeI, and BiTeI. The inset figures show the crystal structures of these materials.}
    \label{fig:SLME}
\end{figure}

In \textbf{Figure 4a}, we compare the predicted and true efficiencies for the test set, in which $R^2 = 0.81$ shows a high prediction accuracy of the GNNOpt for the SLME. Using the trained GNNOpt model, we predict the $\eta$ values of 5,281 unseen crystal structures from the Materials Project without ground-truth optical spectra ~\cite{jain2013commentary}. Here, we select only the stable insulators with an energy band gap $E_g$, suitable for solar-energy harvesting applications ($0 \leq E_g \leq 5.0$ eV). Statistical plots of the unseen materials with the number of materials as a function of $Eg$, $N$, and lattice constant are shown in \textbf{Figure S12} in Supplementary Information. In \textbf{Figure 4b}, we plot the predicted-$\eta$ as a function of $E_g$. We observe that the maximum $\eta$ is about 32\% at $E_g \sim 1.3$ eV, which is consistent with the SQ limit. However, the SLME is a more stringent selection parameter than the SQ limit for the solar cell materials because the SLME shows $\eta$ values over a wide range for materials with similar band gaps, indicating the significant contribution of $\alpha(E)$ to $\eta$. 

Additionally, knowing which elements from the periodic table contribute most to high-efficiency solar cell materials can serve as an initial guideline for material design. As presented in \textbf{Figure 4c}, the transition metals such as Tc, Rh, Pd, Pt, Cu, Ag, Au, and Hg, and chalcogenides including S, Se, and Te, are the main constituent elements of the solar cell materials. This finding is in agreement with widely known solar cell materials, such as Cu-rich chalcopyrite~\cite{jiang2020highly}, Pb-based perovskites~\cite{tai2019recent,ke2019unleaded}, or CdTe~\cite{ferekides2000high}. In \textbf{Table S1} in Supplementary Information, we identified 246 materials with $\eta$-SLME greater than 32\%. To validate the predicted SLME of the unseen materials, we selected three examples from the highest-SLME materials list: LiZnP, SbSeI, and BiTeI. These materials are not present in the DFT database. We conducted DFT calculations to determine $\alpha(E)$ in these examples. The results, depicted in \textbf{Figure 4d}, show excellent agreement between the DFT calculations (dashed lines) and the GNNOpt-predicted $\alpha$ values (solid lines). This indicates that GNNOpt could be an effective materials screening tool at a much lower computational cost.

\section{Probing quantum materials}

\begin{figure}[t]
    \centering
    \includegraphics[width=0.7\linewidth]{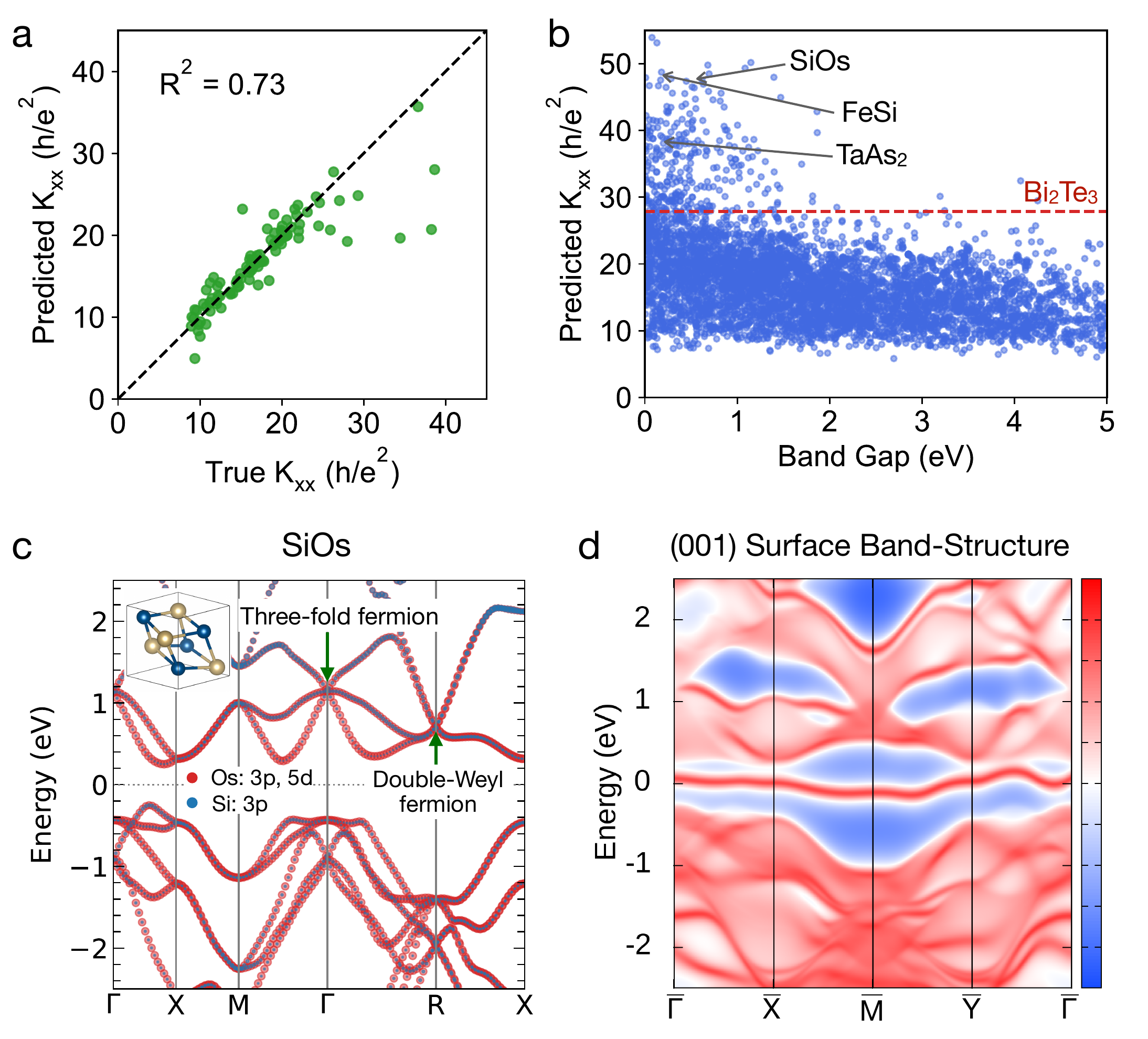}
    \caption{\textbf{Searching for quantum materials with high Quantum weight $K_{xx}$ using GNNOpt.} 
    \textbf{a.} Comparison between predicted and true $K_{xx}$, obtained by the GNNOpt and the DFT ground truth, respectively, for the test set. \textbf{b.} $K_{xx}$-predicted by the GNNOpt of the 5,281 stable materials from the Material Project, which is plotted as a function of the energy band gap. The dashed line indicates $K_{xx}=27.87$ of Bi$_2$Te$_3$. \textbf{c.} Electronic band structure with atomic orbital projections in the color of the high-$K_{xx}$ material, SiOs. The arrows indicate the double-Weyl and three-fold fermions at the $\Gamma$ and $R$ points, respectively. \textbf{d.} The surface band structure along the high-symmetry lines on the (001) surface of SiOs. The color bar scale is in arbitrary units.}
    \label{fig:wq}
\end{figure}

In this section, we provide another application scope of GNNOpt to probe quantum geometry and topology in quantum materials. Since the dipole moment matrix element for optical transition is closely related to the interband Berry connection~\cite{aversa1995nonlinear}, recent studies have established the relationship between the quantum geometry and the optical properties~\cite{ahn2020low,holder2020consequences,bhalla2022resonant}. In particular, very recently, Onishi and Fu~\cite{onishi2024fundamental} theoretically showed that the generalized quantum weight, a concept that can be derived from optical spectra, is a direct measure of ground state quantum geometry and topology. The quantum weight $K_{xx}$ is given by a modification of $f$-sum rule weighted by the inverse frequency as follows~\cite{onishi2024fundamental}:
\begin{equation}
    K_{xx}= \frac{2\hbar}{e^2}\int_0^\infty \frac{\text{Re}[\sigma_{xx}(\omega)]}{\omega}\mathrm{d}\omega=\frac{2\hbar}{e^2}\int_0^\infty \epsilon_2(\omega)\mathrm{d}\omega,
\end{equation}
where $\text{Re}[\sigma_{xx}(\omega)]$ is the real part of optical conductivity over the entire frequency range (i.e., $0<\omega<\infty$).

The GNNOpt for prediction $\epsilon_2(\omega)$ with $0<\omega < 50$ eV (see \textbf{Figure S6}) is used to obtain $K_{xx}$. It is noted that for $\omega > 50$ eV, $\epsilon_2(\omega) \to 0$ for all materials in the training set. In \textbf{Figure 5a}, we compare the predicted and true $K_{xx}$ in the unit of $h/e^2$ for the test set. $R^2 = 0.73$ shows a good prediction between the GNNOpt and the DFT, particularly for $K_{xx} < 25$. The GNNOpt is thus used to predict $K_{xx}$ of the 5,281 unseen insulator materials, as shown in \textbf{Figure 5b}. For simplicity, we consider the quantum weight of the well-known topology insulator Bi$_2$Te$_3$ ($K_{xx} = 28.87$) as a threshold for classifying quantum materials, in which the material with $K_{xx} > 28.87$ is considered as the high-$K_{xx}$ materials. We identified 297 high-$K_{xx}$ materials, listed in \textbf{Table S2} in Supplementary Information, in which several materials, such as ZrTe$_5$ ($K_{xx} = 33.90$), TaAs$_2$ ($K_{xx} = 37.66$), FeSi ($K_{xx} = 48.74$), and NbP ($K_{xx} = 35.58$), etc., have been reported as quantum materials with anomalous hall effect~\cite{liang2018anomalous}, large magnetoresistance~\cite{wang2016resistivity}, topological Fermi arcs~\cite{changdar2020electronic}, and quantum oscillations~\cite{klotz2016quantum}. 

To further validate the quantum characteristics from the predicted high-$K_{xx}$ materials, we carry out additional DFT calculations for SiOs, which have exceedingly high quantum weight ($K_{xx} = 46.52$) but have not been well studied. As shown in \textbf{Figure 5c}, SiOs is found to host the three-fold and double-Weyl fermions at the $\Gamma$ and $R$ points, respectively. Since both the three-fold and double-Weyl fermions are located above the Fermi level ( at 1.15 eV and 0.7 eV, respectively), it would not be possible to detect them using the angular resolved photoemission spectroscopy (ARPES) technique. Nevertheless, irrespective of the energy position of the manyfold band crossing points, we could expect the associated Fermi surface arcs on the surface Brillouin zone~\cite{bradlyn2016beyond,sun2015prediction}. We thus calculated the surface band structure of the (001) SiOs by using the maximally-localized Wannier functions and the Green's function approach (see Method). As shown in \textbf{Figure 5d}, we find the surface state near the Fermi level form self-enclosed loops, indicating the triviality of these states, in contrast to the open Fermi arcs. This closed-loop shrinks in size when the surface state moves away from the Fermi energy (see \textbf{Figure S15}a). However, the nontrivial topological Fermi arcs associated with manyfold fermions can be found at 1.3 eV above the Fermi level (see Figure S15b). These signatures suggest the ultra-quantum properties of SiOs.

\section{Discussions}
In this work, we present GNNOpt, a graph neural network model with engineered input embedding that directly predicts all linear optical spectra from crystal structures. The ensemble embedding layer for automatic embedding optimization improves prediction accuracy without modifying the neural network structures. By integrating with the equivariant neural network, GNNOpt achieves high-quality predictions with high data efficiency using a small training set of 944 materials. Furthermore, applying the Kramers-Kr{\"o}nig relation before training the GNNOpt model results in better-predicted optical spectra, as observed by comparing optical properties before and after training. With GNNOpt, we were able to identify over 200 materials with over 32\% conversion efficiency for solar energy harvesting applications. Additionally, thanks to the recent connection between optical property and ground-state topology, we were able to screen quantum materials carrying multiple nontrivial topology, including SiOs. 

The direct ensemble structure input in GNNOpt opens several promising future research directions. First, the atomic embedding with multiple features greatly facilitates the encoding of defect structures, which could be encoded by perturbing one or more atomic features. Given the significant impact of defects on optical properties, predicting optical properties with defects is highly desirable. This approach could enable tuning absorption spectra through new defect levels or identifying non-radiative recombination centers that reduce solar cell efficiency. Second, building on the successful prediction of linear optical properties, extending this method to second and third-order optical responses, such as second harmonic generation and Raman spectroscopy, offers valuable opportunities. Third, since the current ground truth database is DFT-based, future studies incorporating many-body effects, such as the GW approximation, could provide more accurate predictions, particularly for excitons. Lastly, since the ensemble embedding integrates multiple embedding types to enhance overall model performance and goes beyond feature selection, it is expected to be broadly applicable to other GNN models.

\section{Method}
\subsection{Data preparation for optical spectra}
We trained and tested the GNNOptic with the 944 crystalline solids calculated from the density functional theory (DFT) by Yang \text{et al.}~\cite{yang2022high}. The databases are obtained from Materials Project~\cite{jain2013commentary} using the API. The optical spectra in the database include the frequency-dependent dielectric function and the corresponding absorption coefficient, which was performed within the independent-particle approximation (IPA). Yang \text{et al.}~\cite{yang2022high} showed that the IPA is sufficient in most cases to reproduce the experiment spectra. 
The 940 materials in the database were selected based on the energy band gap between $0 \leq E_g \leq 5.0$ eV and the atomic site number $N < 10$ in each unit cell. The static plot of the number of materials as a function of $E_g$, $N$, and lattice constants are given in Figure S1 in Supplementary Information.

We randomly split the entire dataset into 80\%, 10\%, and 10\% for the training (733 materials), validation (97 materials), and test (110 materials) sets, respectively (see Figure S2 in Supplementary Information). Given limited training data, the high photon energy $\omega$ resolution (with 2001 points) requests a large number of parameters, such as radius cut-off and the number of layers, in the neural network to fit the output dimension. Therefore, to ensure a balanced output dimension and the calculation cost, we interpolate the smoothed spectrum in the range $0 \leq \hbar\omega \leq 50$ eV to 251 points, which is suitable to reproduce all of the spectra peaks (see Figure S3 in Supplementary Information). We also apply the predictive model on 5,281 stable insulators from the Materials Project~\cite{jain2013commentary} with $0 \leq E_g \leq 5.0$ eV with atomic site number $N < 20$ in each unit cell. The static plot of the number of materials as a function of $E_g$, N, and lattice constants for this database are given in Figure S12 in Supplementary Information.

\subsection{Hyperparameters optimization}
The set of optimized parameters to get the best results for GNNOpt is given as follows: the maximum cutoff radius $r_\text{cut}= 6$ \AA\, the maximum of spherical harmonics $l_\text{max}=2$, the length of the embedding feature vector is 64, the multiplicity of irreducible representation is 32, the number of pointwise convolution layer $n_\text{conv}=2$, the AdamW optimizer learning rate is $(5\times 10^{-3})\times 0.96^{k}$ with $k$ being the epoch number, and the AdamW optimizer weight decay coefficient is 0.05.

\subsection{First-principles calculations}
We use the Quantum ESPRESSO package for the DFT calculations to obtain the absorption coefficient $\alpha(E)$ of LiZnP, SbSeI, and BiTeI within independent particle approximation~\cite{giannozzi2009quantum,hung2022quantum}. The crystal structures of these materials are taken from the Material Project to be consistent with that in the GNNOpt prediction. The optimized norm-conserving Vanderbilt pseudopotentials~\cite{hamann2013optimized} with the Perdew–Burke–Ernzerhof exchange-correlation functional~\cite{perdew1996generalized} are selected for all atoms. A cutoff energy of 60 Ry for plane wave is used for all materials. For self-consistent field (SCF) calculations, the k-points meshes are $6\times 6 \times 6$, $4\times 8 \times 3$, and $9\times 9 \times 5$ for LiZnP, SbSeI, and BiTeI, respectively, while the dense meshes of $20\times 20 \times 20$, $14\times 30 \times 12$, and $27\times 27 \times 15$, respectively, are used for non-SCF calculations in order to achieve convergence in $\alpha(E)$.

To calculate the electronic band structure of SiOs, the tight-binding model is obtained by the maximally-localised Wannier functions using the Wannier90 code~\cite{mostofi2008wannier90}, in which Si $3p$ and Os $3p, 5d$ orbitals are selected as the basis. Then, the WannierTools code~\cite{wu2018wanniertools} with the Green function approach is used for the analysis of the surface band structure.

\medskip
\textbf{Supporting Information} \par 
Supporting Information is available from the Wiley Online Library.

\medskip
\textbf{Acknowledgements} \par 
N.T.H. acknowledges the Researcher, Young Leaders Overseas Program and financial support from the Frontier Research Institute for Interdisciplinary Sciences from Tohoku University. R.O. acknowledges the support from the U.S. Department of Energy (DOE),  Office of Science (SC), Basic Energy Sciences (BES), Award No. DE-SC0021940. A.C. thanks National Science Foundation (NSF) Designing Materials to Revolutionize and Engineer our Future (DMREF) Program with Award No. DMR-2118448. M.L. is partially supported by NSF ITE-2345084, the Class of 1947 Career Development Chair and the support from R. Wachnik. 

\medskip
\textbf{Conflict of Interest} \par 
The authors declare no conflict of interest.

\medskip
\textbf{Data Availability Statement} \par 
The data that support the findings of this study are openly available on GitHub at \url{https://github.com/nguyen-group/GNNOpt}.

\medskip

%
\bibliographystyle{MSP}



\end{document}


\newpage
\flushbottom
\tableofcontents
\thispagestyle{empty}
\section{Dataset for training GNNOpt model}
\begin{figure}[ht]
    \centering
    \includegraphics[width=0.95\linewidth]{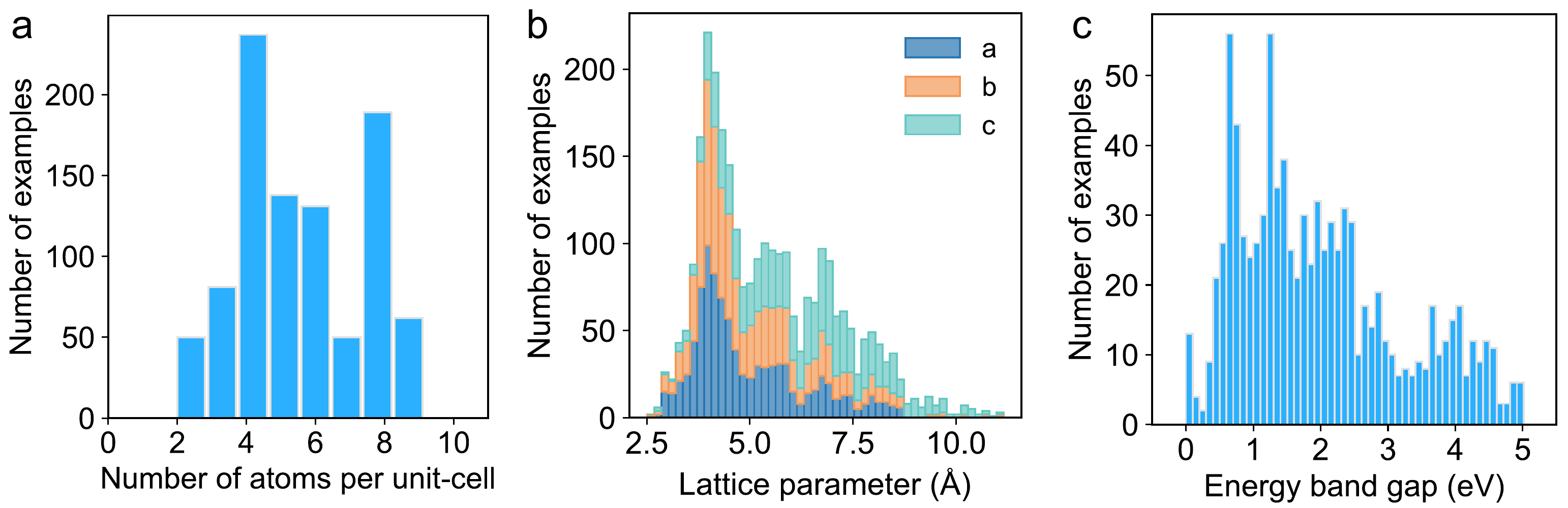}
    \caption{The distribution of (a) the number of atoms per unit cell, (b) the lattice parameters, and (c) the energy band gap. The total 944 materials from the Materials Project contain 2 to 9 atoms per unit cell and the energy band gap from 0 to 5 eV.}
    \label{fig:S1}
\end{figure}

\begin{figure}[ht]
    \centering
    \includegraphics[width=1.0\linewidth]{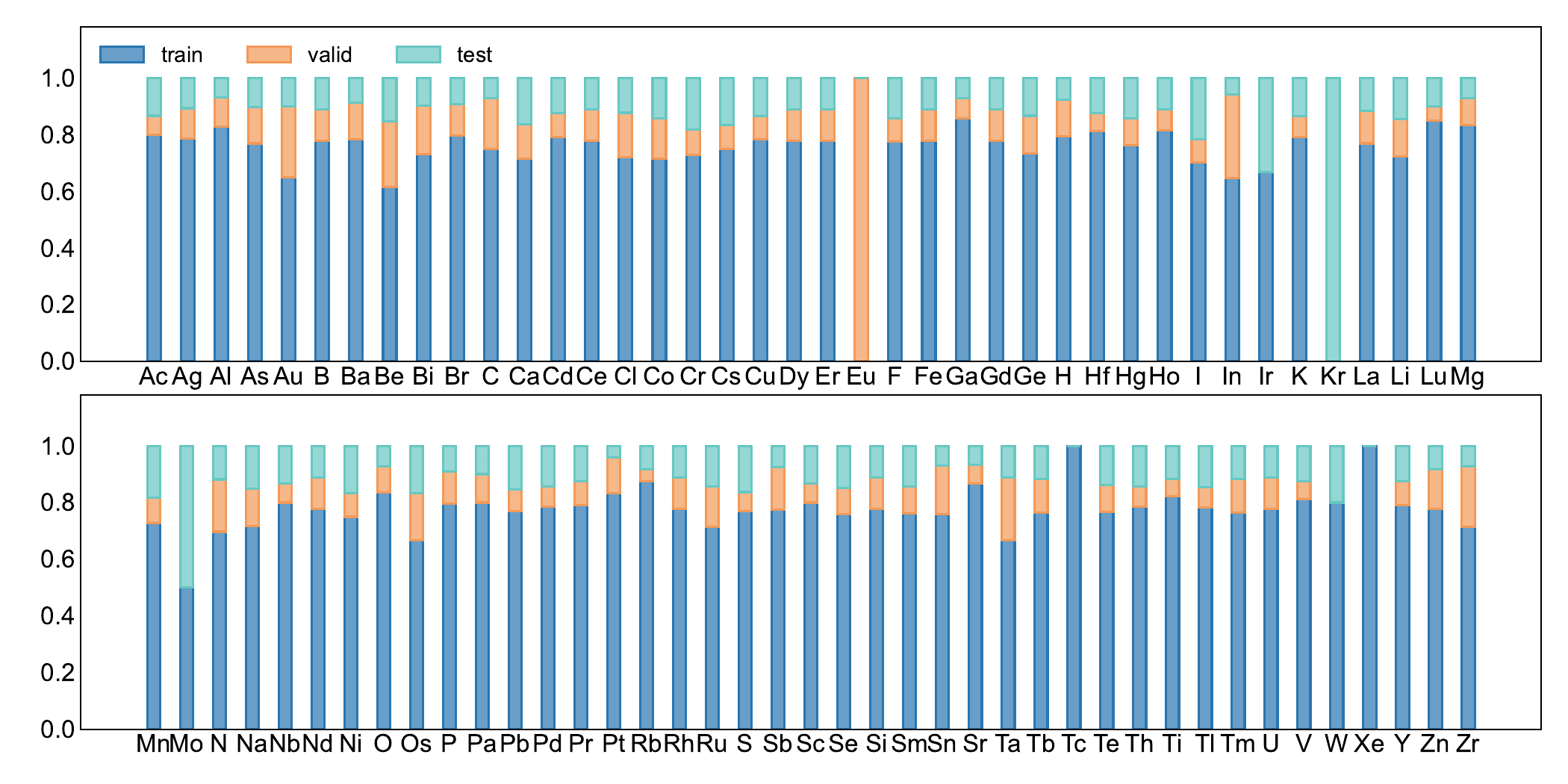}
    \caption{The distribution of the training, validation, and testing data by elements, in which the data is randomly split into 80\%, 10\%, and 10\%, respectively.}
    \label{fig:S2}
\end{figure}

\begin{figure}[ht]
    \centering
    \includegraphics[width=0.65\linewidth]{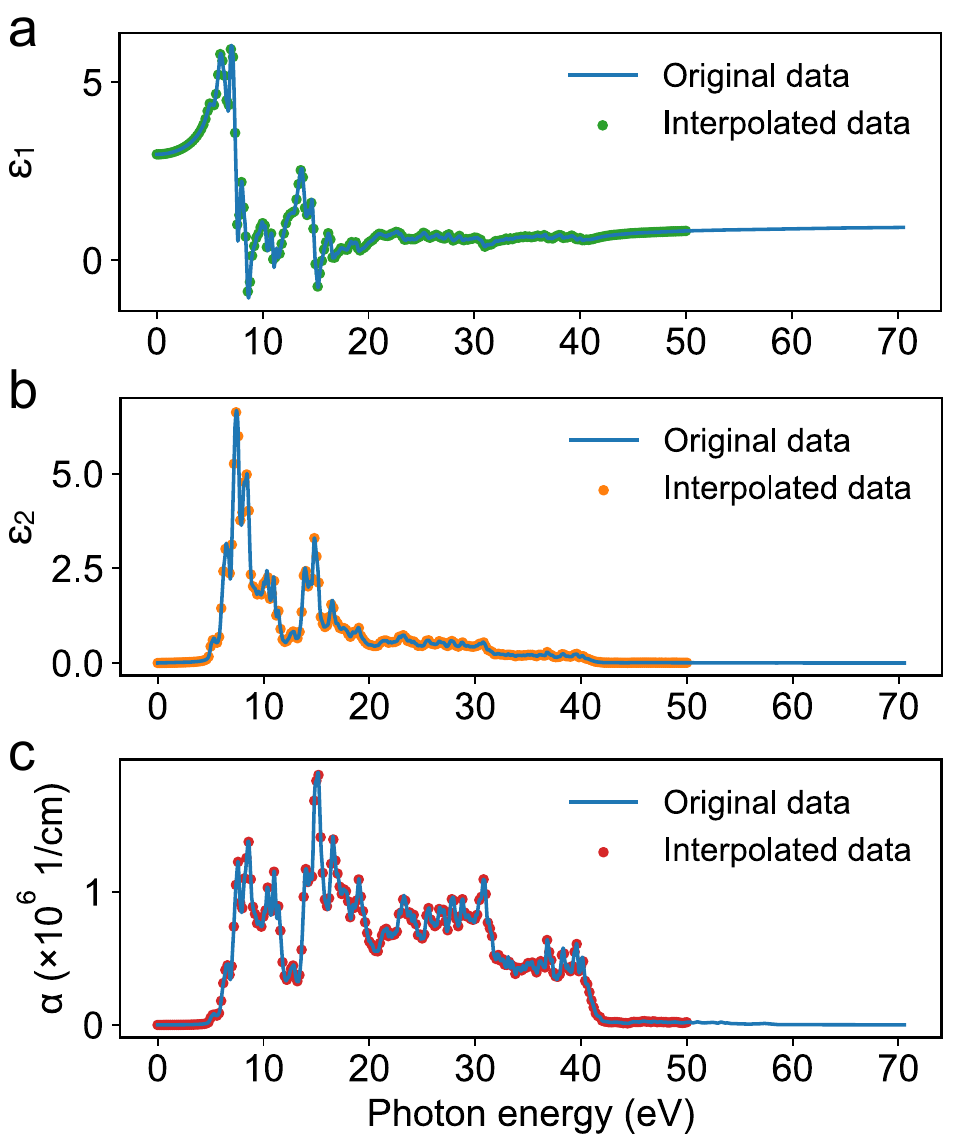}
    \caption{Interpolation data with 251 points ranging from 0 to 50 eV from the original DFT spectra (solid lines). (a) The real $\epsilon_1$, (b) imaginary $\epsilon_2$ parts of the dielectric function, and (c) the absorption coefficient $\alpha$ are plotted as a function of the photon energy.}
    \label{fig:S3}
\end{figure}

The data set for training the GNNOpt model includes 944 materials obtained from the Materials Project. The distribution of the number of atoms per unit cell, lattice parameters, and energy band gap is plotted in Figure~\ref{fig:S1}. The distribution of the training, validation, and testing datasets is plotted in Figure~\ref{fig:S2}. In Figure~\ref{fig:S3}, the original DFT data is interpolated into 251 points ranging from 0 to 50 eV by using NumPy.linspace() function. 

\clearpage
\section{Performance of GNNOpt for optical prediction}
\subsection{Accuracy of model}
\begin{figure}[ht]
    \centering
    \includegraphics[width=0.95\linewidth]{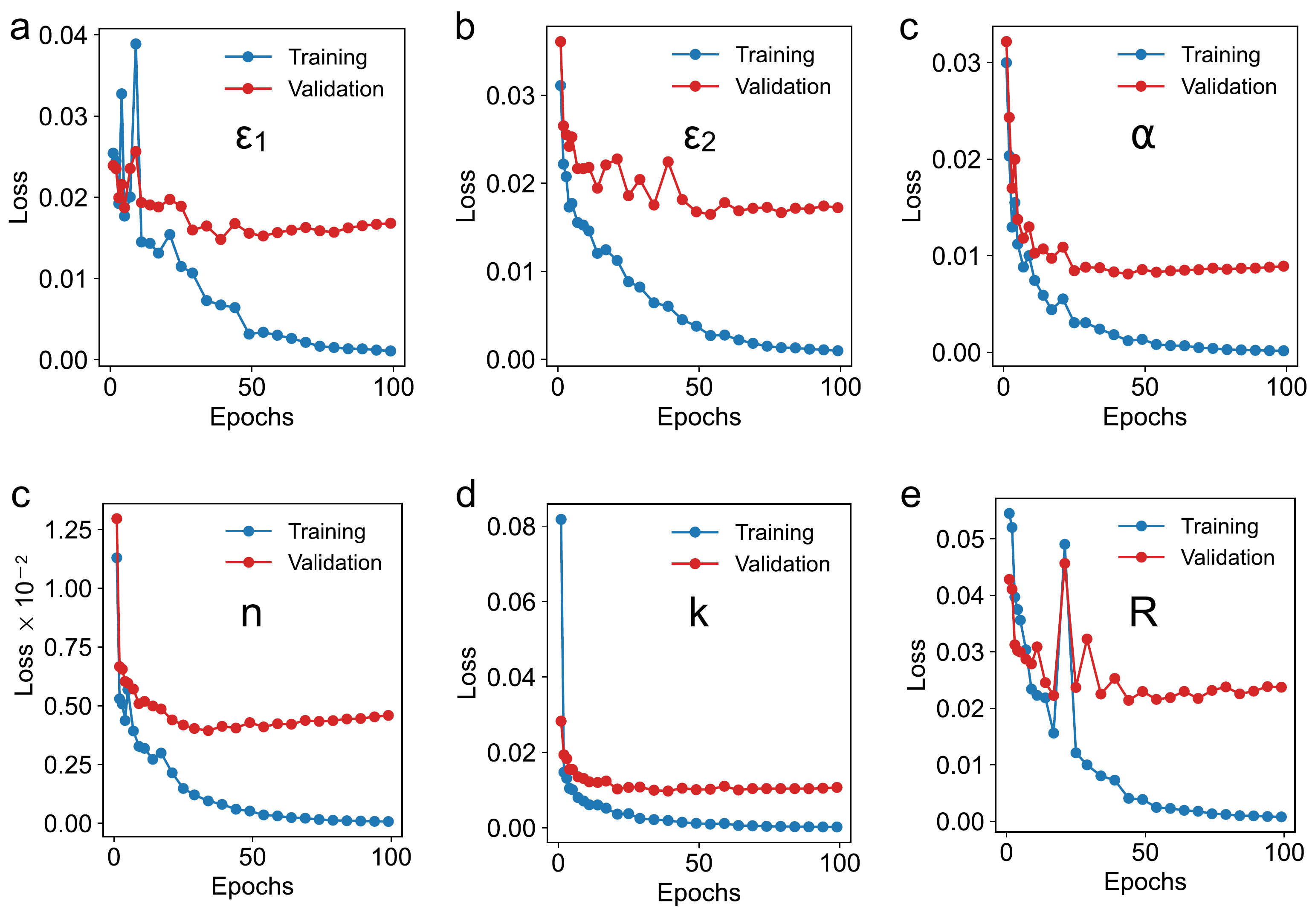}
    \caption{Loss history as a function of epoch number for the (a) real $\epsilon_1$ and (b) imaginary $\epsilon_2$ parts of the dielectric function, (c) absorption coefficient $\alpha$, (d) real $n$ and (e) imaginary $k$ parts of refractive index, and (f) reflectance $R$.}
    \label{fig:S4}
\end{figure}

\begin{figure}
    \centering
    \includegraphics[width=0.95\linewidth]{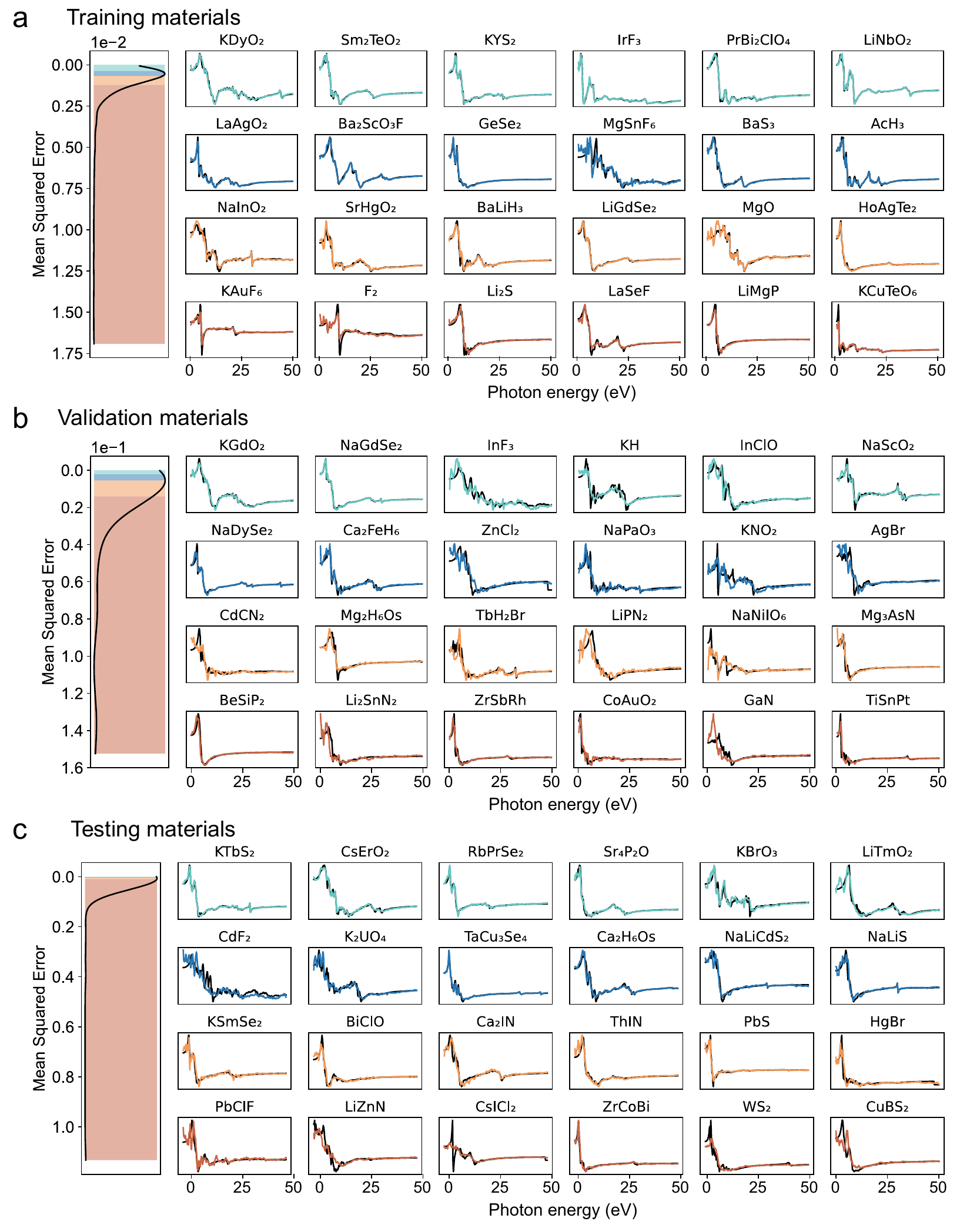}
    \caption{Direct prediction results for real $\epsilon_1$ part of the dielectric function of training, validation, and testing datasets. Color lines are the GNNOpt-predicted spectra, and black lines are the DFT ground-truth spectra. 24 random materials are selected for each dataset, and the selected materials correspond to each error quartile in the mean squared error distribution (left figure) with the same color.}
    \label{fig:S5}
\end{figure}

\begin{figure}
    \centering
    \includegraphics[width=0.95\linewidth]{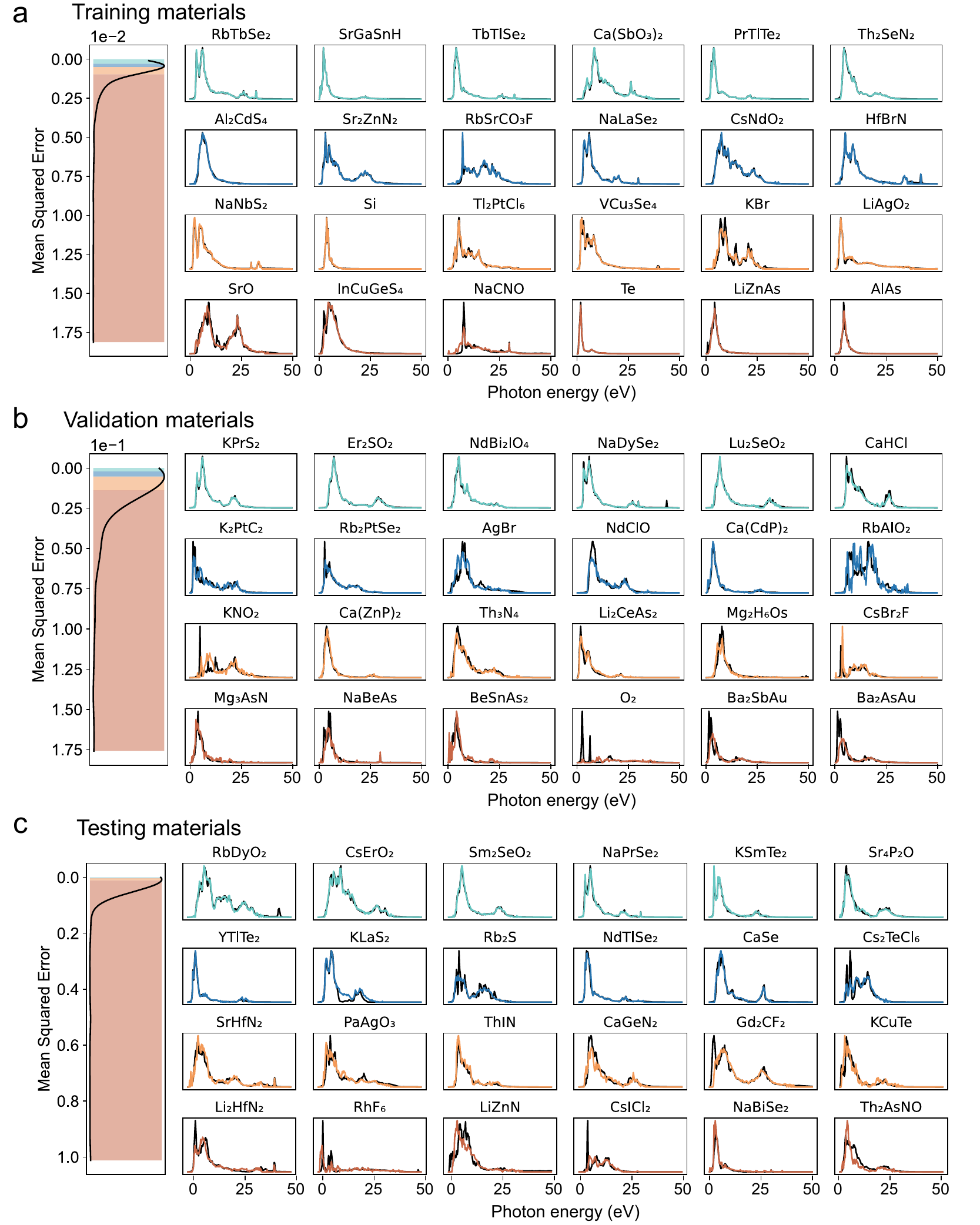}
    \caption{Direct prediction results for imaginary $\epsilon_2$ part of the dielectric function of training, validation, and testing datasets. Color lines are the GNNOpt-predicted spectra, and black lines are the DFT ground-truth spectra. 24 random materials are selected for each dataset, and the selected materials correspond to each error quartile in the mean squared error distribution (left figure) with the same color.}
    \label{fig:S6}
\end{figure}

\begin{figure}
    \centering
    \includegraphics[width=0.95\linewidth]{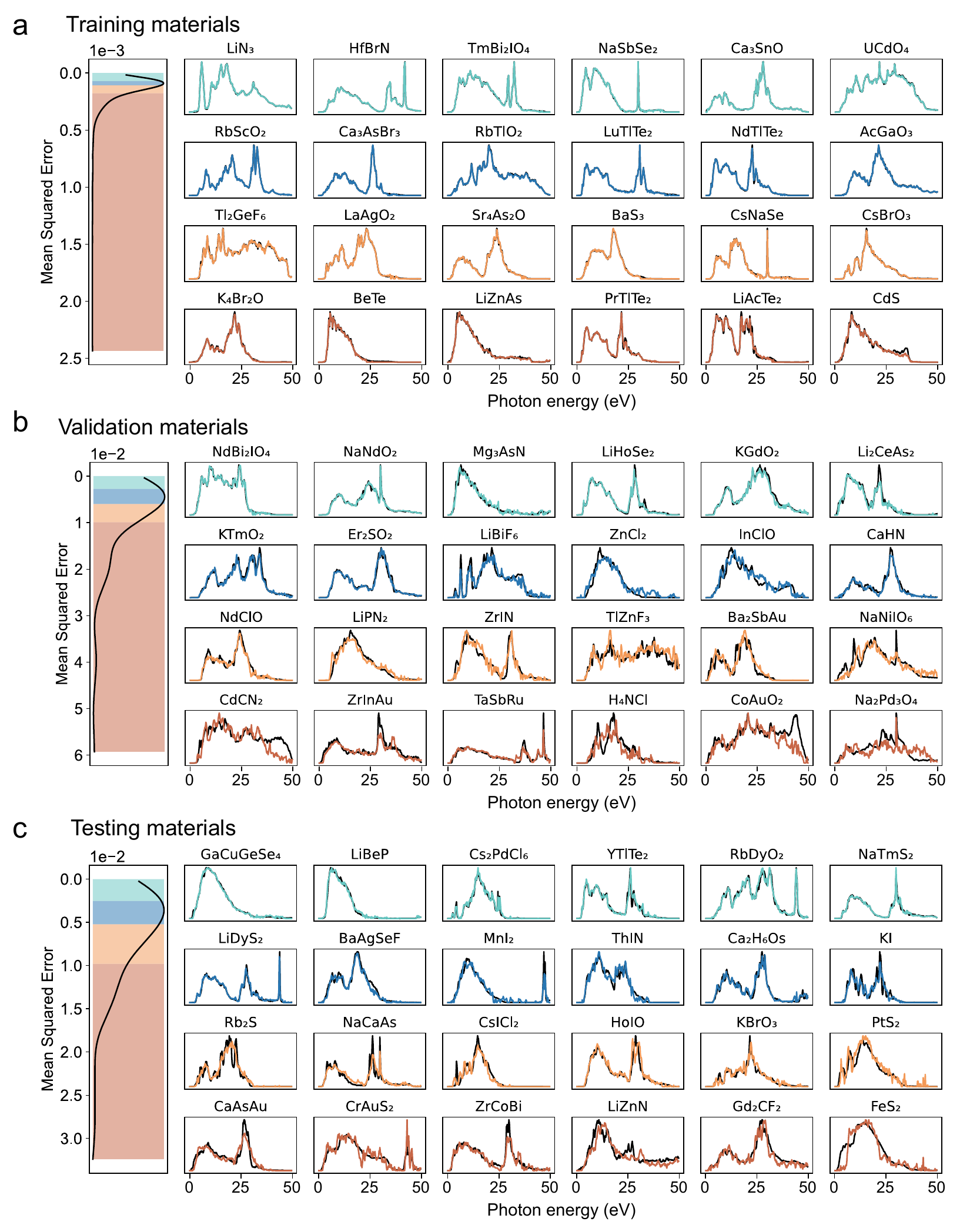}
    \caption{Direct prediction results for absorption coefficient $\alpha$ of training, validation, and testing datasets. Color lines are the GNNOpt-predicted spectra, and black lines are the DFT ground-truth spectra. 24 random materials are selected for each dataset, and the selected materials correspond to each error quartile in the mean squared error distribution (left figure) with the same color.}
    \label{fig:S7}
\end{figure}

\begin{figure}
    \centering
    \includegraphics[width=0.95\linewidth]{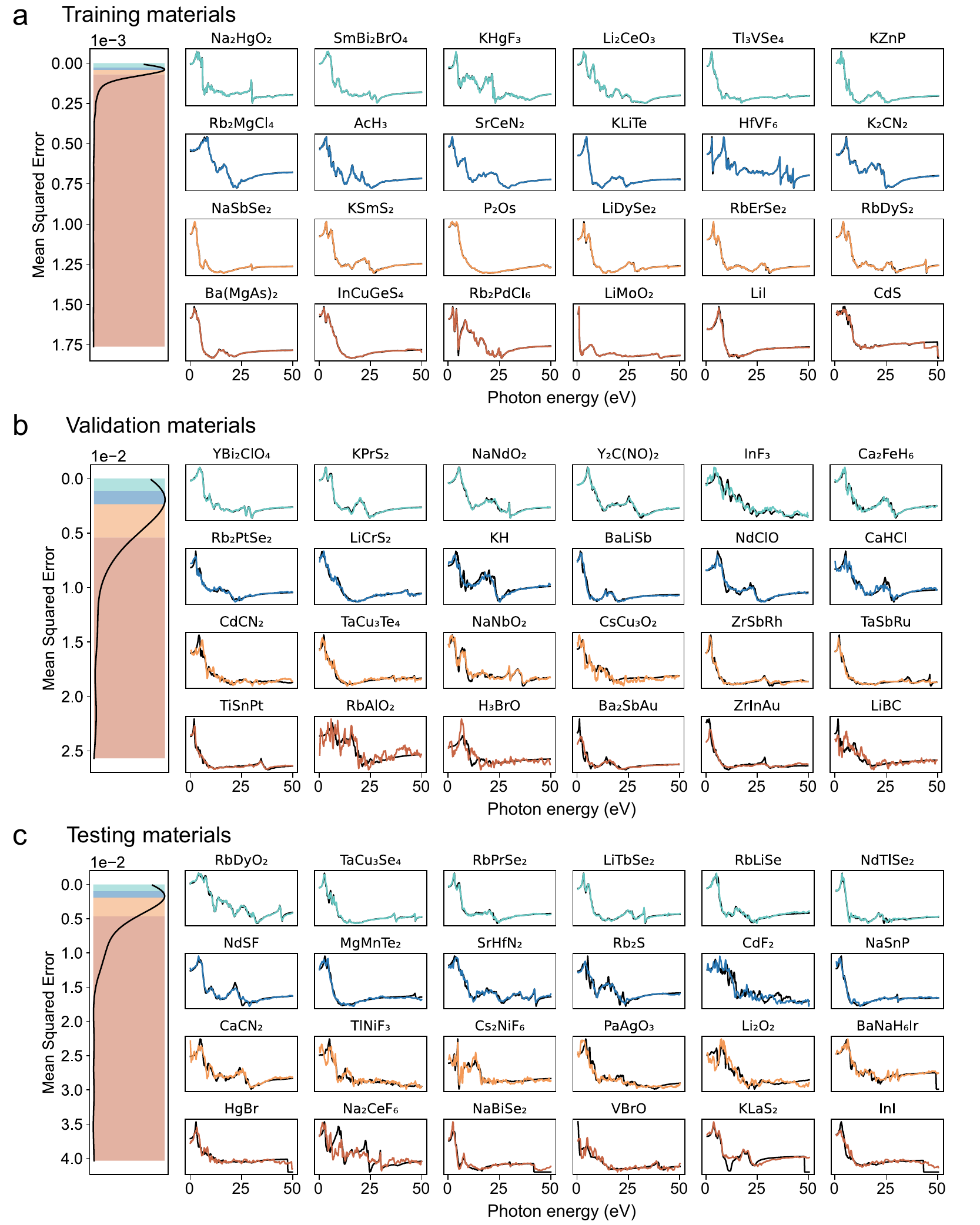}
    \caption{Direct prediction results for real $n$ part of the refractive index of training, validation, and testing datasets. Color lines are the GNNOpt-predicted spectra, and black lines are the DFT ground-truth spectra. 24 random materials are selected for each dataset, and the selected materials correspond to each error quartile in the mean squared error distribution (left figure) with the same color.}
    \label{fig:S8}
\end{figure}

\begin{figure}
    \centering
    \includegraphics[width=0.95\linewidth]{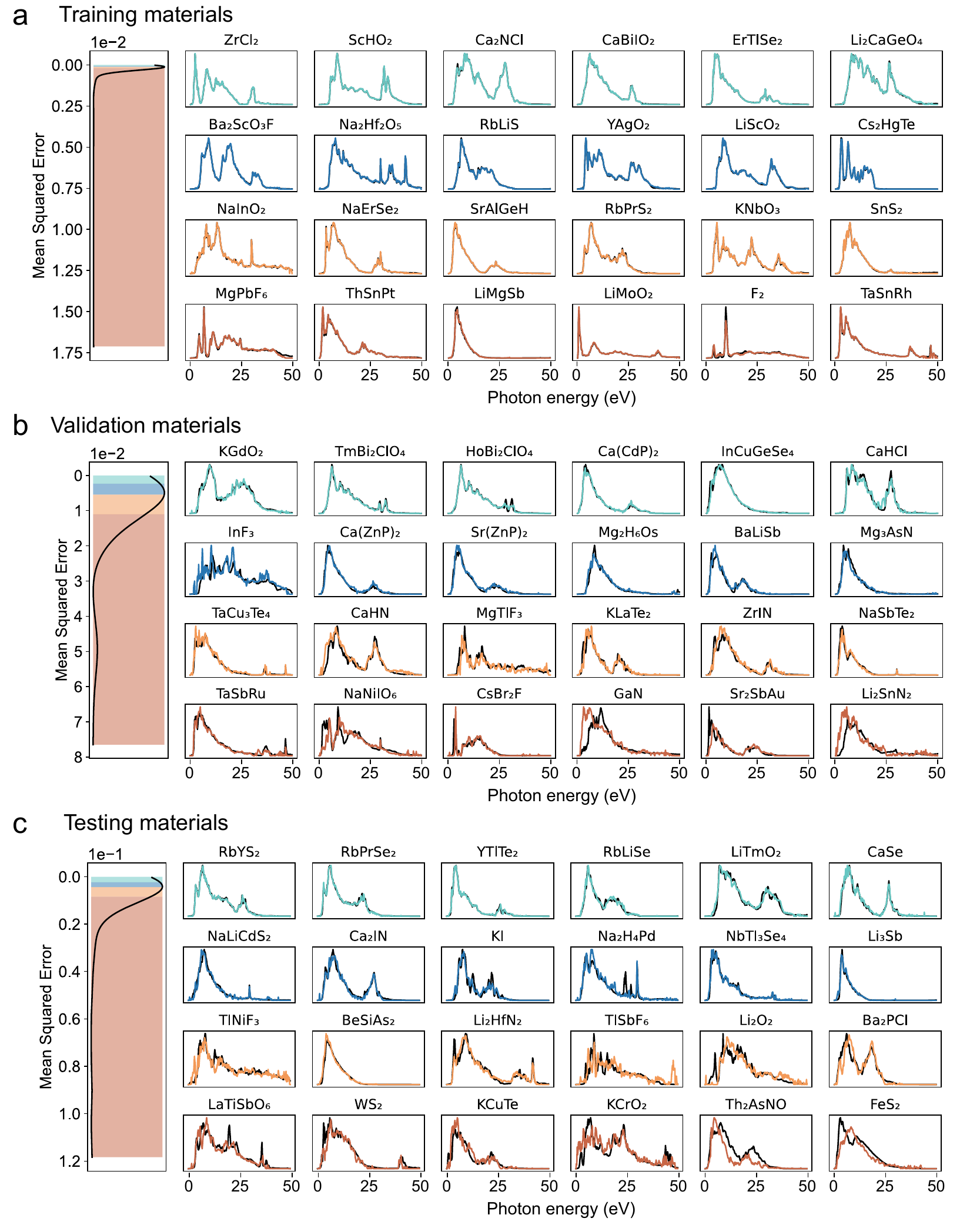}
    \caption{Direct prediction results for imaginary $k$ part of the refractive index of training, validation, and testing datasets. Color lines are the GNNOpt-predicted spectra, and black lines are the DFT ground-truth spectra. 24 random materials are selected for each dataset, and the selected materials correspond to each error quartile in the mean squared error distribution (left figure) with the same color.}
    \label{fig:S9}
\end{figure}

\begin{figure}
    \centering
    \includegraphics[width=0.95\linewidth]{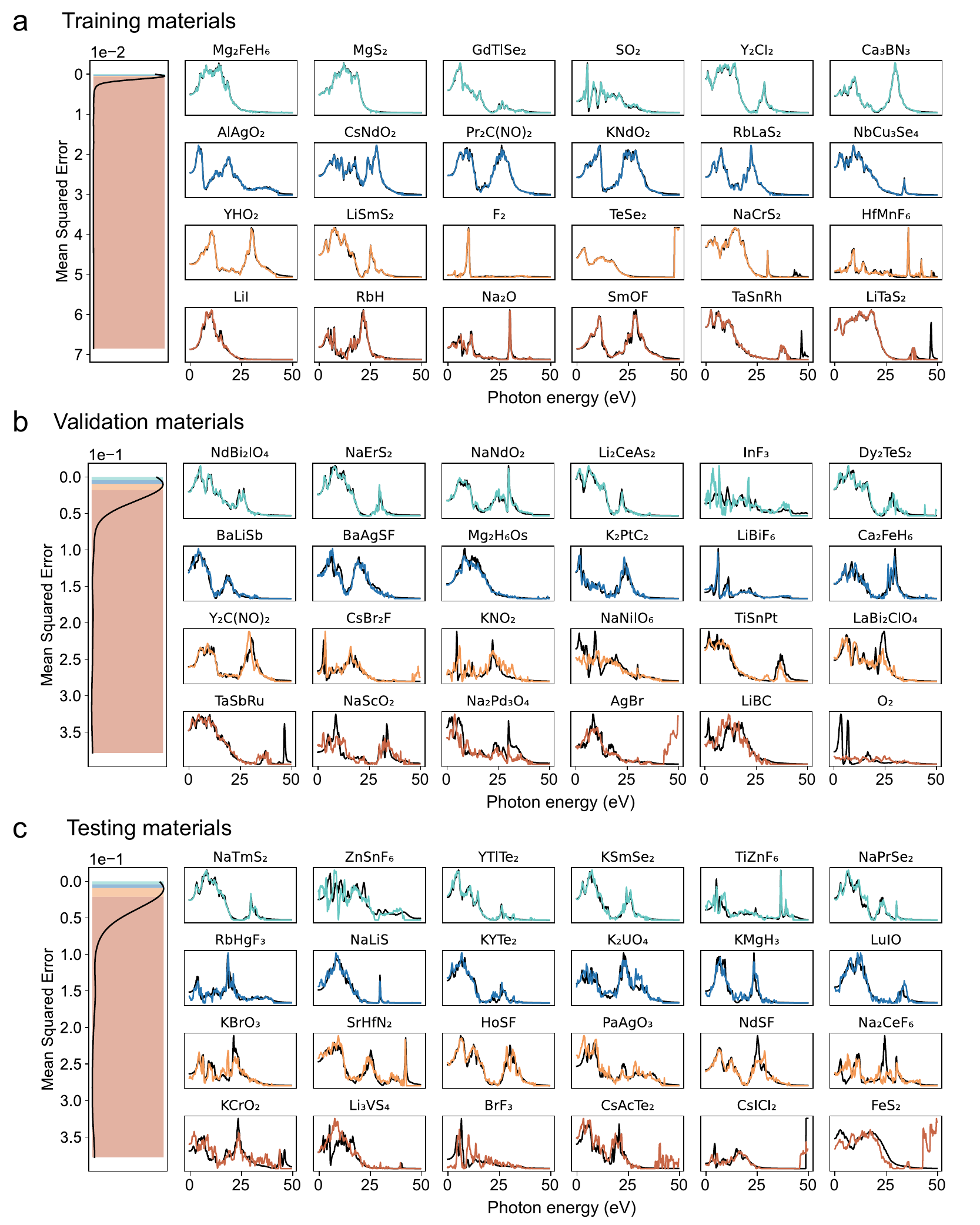}
    \caption{Direct prediction results for reflectance $R$ of training, validation, and testing datasets. Color lines are the GNNOpt-predicted spectra, and black lines are the DFT ground-truth spectra. 24 random materials are selected for each dataset, and the selected materials correspond to each error quartile in the mean squared error distribution (left figure) with the same color.}
    \label{fig:S10}
\end{figure}

We show the loss history of the trained GNNOpt model as a function of epoch number in Figure~\ref{fig:S4} and the direct optical prediction of training, validation, and testing
datasets in Figures~\ref{fig:S5}-\ref{fig:S10} for the real $\epsilon_1$ and imaginary $\epsilon_2$ parts of the dielectric function, absorption coefficient $\alpha$, real $n$ and imaginary $k$ parts of refractive index, and reflectance $R$, respectively.

\clearpage
\subsection{Scalability of model}
\begin{figure}[ht]
    \centering
    \includegraphics[width=0.80\linewidth]{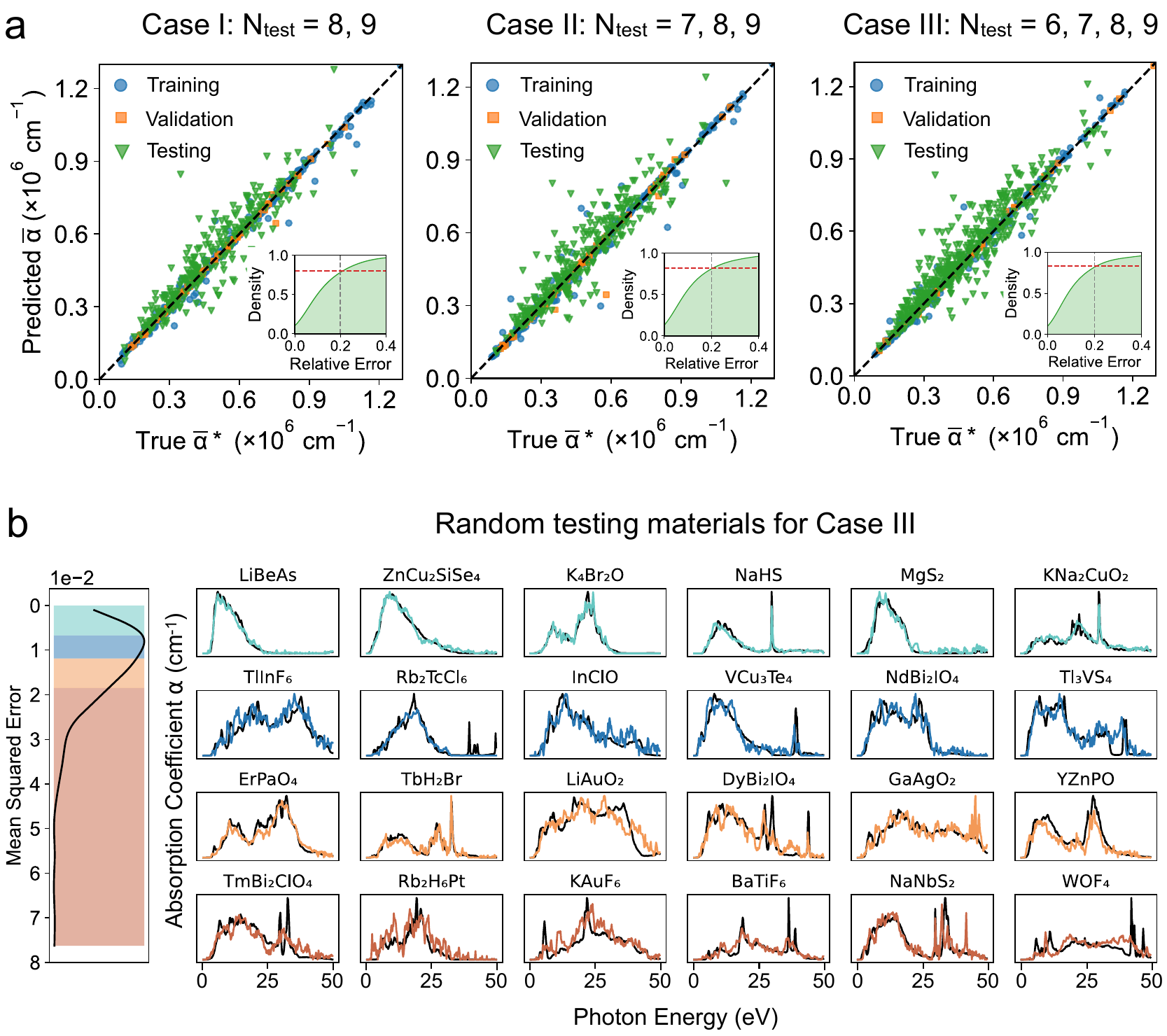}
    \caption{(a) Correlation plots for three cases (I, II, III), which correspond to three GNNOpt models with the testing dataset, including the atomic site numbers $N_\text{test} = 8, 9$, $N_\text{test} = 7, 8, 9$, and $N_\text{test} = 6,7, 8, 9$, respectively. The circle, square, and triangle symbols denote the training, validation, and testing datasets, respectively. The inset figures show the distribution of the relative error. (b) Direct prediction results for absorption coefficient $\alpha$ of a testing dataset of case III. Color lines are the GNNOpt-predicted spectra, and black lines are the DFT ground-truth spectra. 24 random materials are selected for each dataset, and the selected materials correspond to each error quartile in the mean squared error distribution (left figure) with the same color.}
    \label{fig:S11}
\end{figure}

Figure~\ref{fig:S11} shows the performance of the GNNOpt for three cases with the different testing sets, including the atomic site numbers $N_\text{test} = 8, 9$ (case I), $N_\text{test} = 7, 8, 9$ (case II), and $N_\text{test} = 6,7, 8, 9$ (case III). Since $N_\text{train} < N_\text{test}$ for three cases, the GNNOpt shows the scalability to predict the unseen materials, which have a larger number of atoms per unit cell than the training dataset.

\clearpage
\section{Dataset for unseen materials}
\begin{figure}
    \centering
    \includegraphics[width=0.95\linewidth]{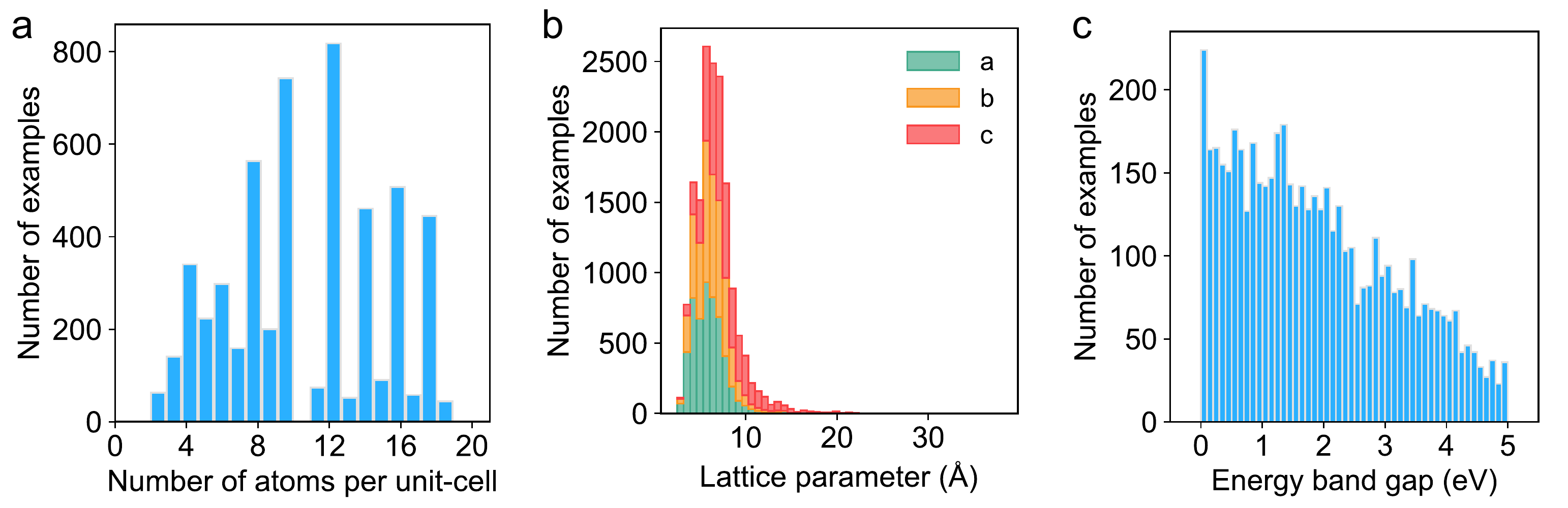}
    \caption{The distribution of (a) the number of atoms per unit cell, (b) the lattice parameters, and (c) the energy band gap. The total 5,281 stable materials from the Materials Project contain 2 to 19 atoms per unit cell and the energy band gap from 0 to 5 eV.}
    \label{fig:S12}
\end{figure}

The data set for unseen materials is obtained from the Materials Project, which includes 5,281 stable insulators (i.e., energy above hull $\geq 0$ eV/atom) and contains 2 to 19 atoms per unit cell and the energy band gap from 0 to 5 eV. The distribution of the number of atoms per unit cell, lattice parameters, and energy band gap is plotted in Figure~\ref{fig:S12}.

\clearpage
\section{GNNOpt for $\alpha(\omega)$ with $0< \hbar \omega < 10$ eV}
\begin{figure}
    \centering
    \includegraphics[width=0.8\linewidth]{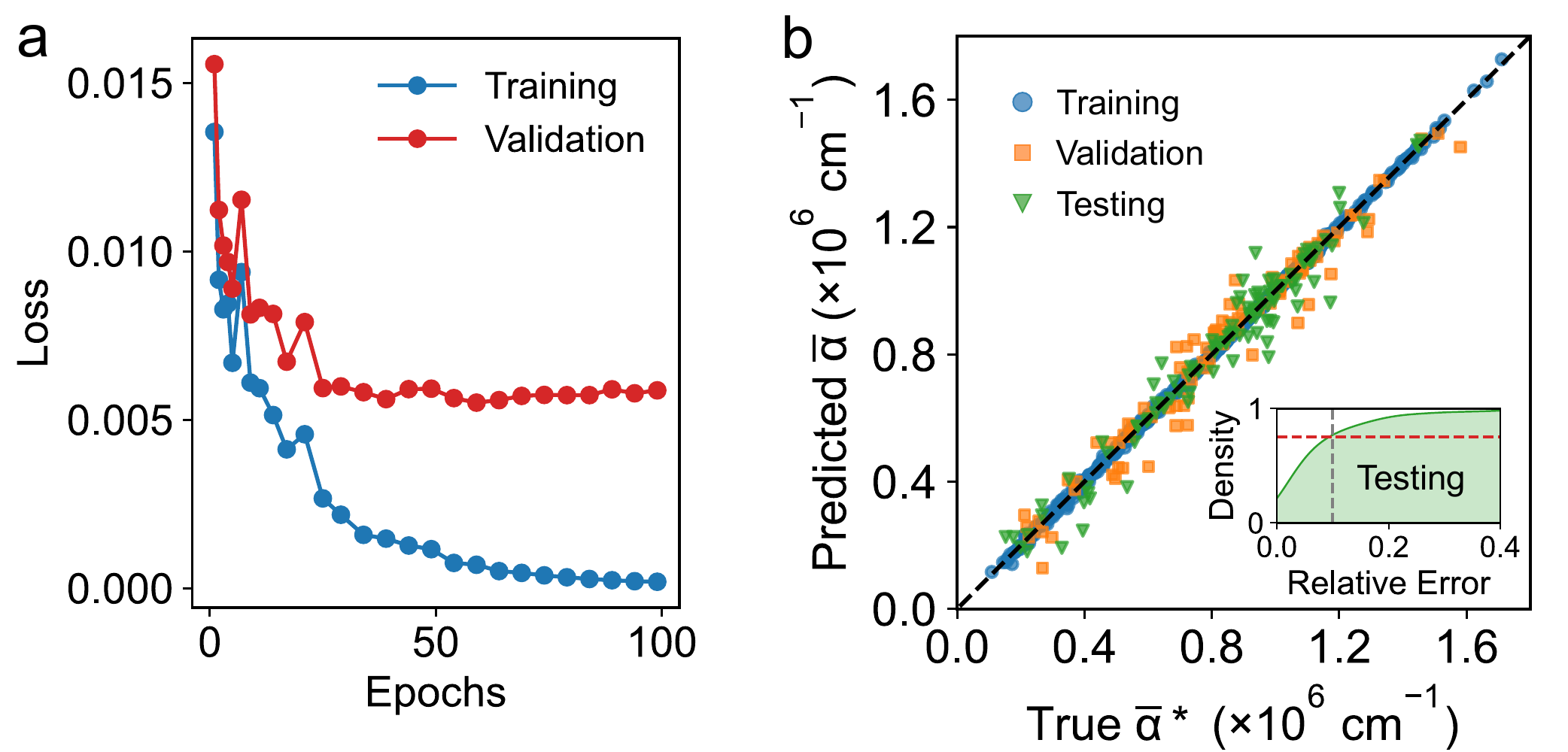}
    \caption{(a) Loss history as a function of epoch number, and (b) correlation plots of training (blue circle), validation (square orange), and testing (triangle green) datasets. The inset figures show the distribution of the relative error.}
    \label{fig:S13}
\end{figure}

In order to increase the resolution of the optical spectra ranging from 0 to 10 eV for the solar cell application. We retrain the GNNOpt model with $0< \hbar \omega < 10$ eV. In Figures~\ref{fig:S13}a and b, we show the loss history and the correlation between the GNNopt-predicted and DFT ground-truth, respectively. $R^2 = 0.95$ for the testing set, and the relative error below for the testing set 10\% is 78\%. 

\begin{figure}
    \centering
    \includegraphics[width=0.95\linewidth]{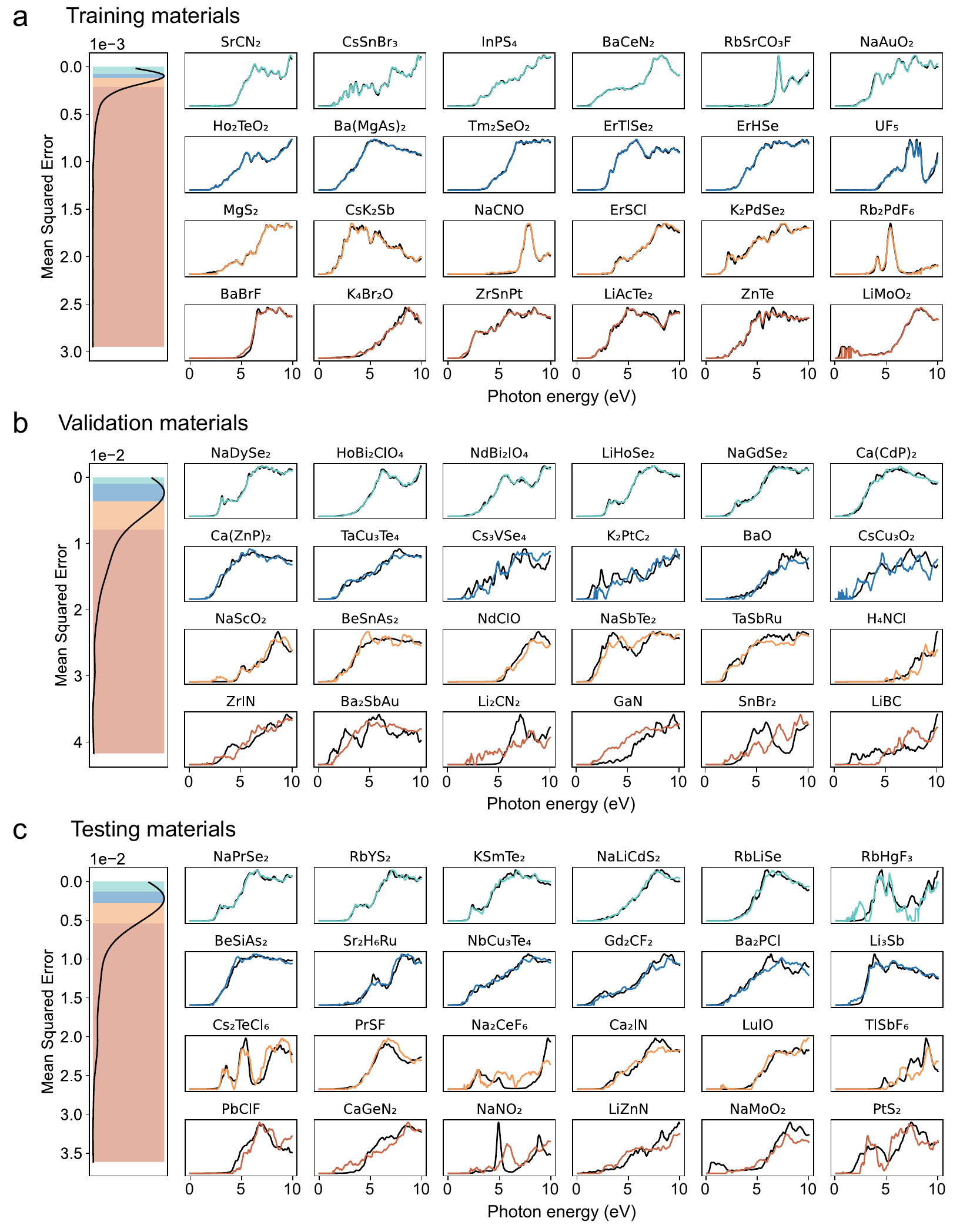}
    \caption{Direct prediction results for absorption coefficient $\alpha$ of training, validation, and testing datasets. Color lines are the GNNOpt-predicted spectra, and black lines are the DFT ground-truth spectra. 24 random materials are selected for each dataset, and the selected materials correspond to each error quartile in the mean squared error distribution (left figure) with the same color.}
    \label{fig:S14}
\end{figure}

\clearpage
\section{Surface state Fermi map of SiOs}
\begin{figure}[ht]
    \centering
    \includegraphics[width=0.95\linewidth]{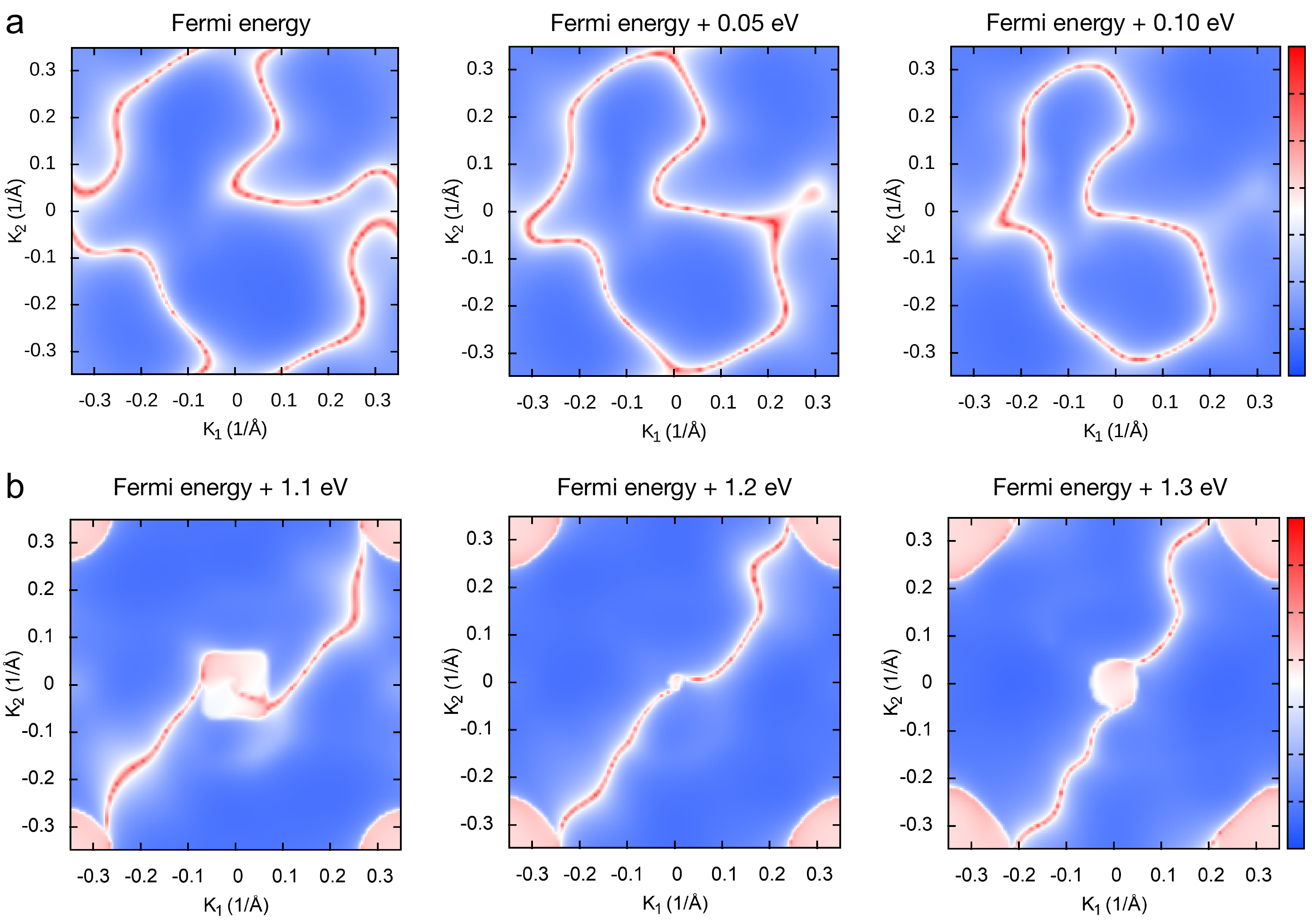}
    \caption{Surface state Fermi map of SiOs taken at (a) Fermi energy $E_F$, $E_F+0.05$ eV, $E_F+0.1$ eV and (b) $E_F+1.10$, $E_F+1.2$ eV, and $E_F+0.13$ eV. The color bar scale is in arbitrary units. The trivial surface is found near the Fermi level with the closed loops, while the nontrivial topological surface is located around $E_F+1.2$ eV with the Fermi arcs.}
    \label{fig:S15}
\end{figure}

In order to check the trivial or nontrivial surface state on the (001) surface of SiOs, we plot the Fermi maps in Figure~\ref{fig:S15}.

\clearpage
\section{Tables of the candidates}
Tables~\ref{tab:1} and~\ref{tab:2} list the material candidates for the solar cell and the quantum materials, respectively. By using the MP-ID column, one can find the detailed structure from the Materials Project website. In Table~\ref{tab:1}, we list only the materials with the solar cell efficiency $\eta > 32$\%, in which $\eta$ is evaluated by using the spectroscopic limited maximum
efficiency (SLME) approach. In Table~\ref{tab:1}, we list only the materials with the quantum weight $K_{xx} > 28.87$, in which $K_{xx} = 28.87$ is the quantum weight of the well-known topology insulator Bi$_2$Te$_3$.

\DefTblrTemplate{middlehead,lasthead}{default}{Continued from previous page}

\subsection{High-SMLE materials for solar cell}
{\small
\begin{longtblr}
[
caption = {The list of 246 high-$\eta$ materials with the Materials Project IDs (MP-ID). The solar cell efficiency $\eta$ (the spectroscopic limited maximum eﬃciency) in the unit of \% and the energy band gap $E_g$ in the unit of eV.},
label = {tab:1}
]
{
colspec={|l|l|l|l||l|l|l|l|},
hline{1,Z}={1}{-}{0.05em},
hline{1,Z}={2}{-}{0.05em},
hline{2}={0.05em}
}
MP-ID & Formula & $E_g$ & $\eta$ & MP-ID & Formula & $E_g$ & $\eta$ \\
        mp-861942 & Ag2GePbS4 & 1.370 & 32.918 & mp-7439 & K3Cu3P2 & 1.289 & 32.251 \\ 
        mp-556866 & Ag2HgSI2 & 1.221 & 32.389 & mp-9273 & K3Sb2Au3 & 1.105 & 32.155 \\ 
        mp-36216 & Ag2S & 1.296 & 32.347 & mp-8704 & K3SbSe4 & 1.432 & 32.275 \\ 
        mp-29163 & Ag2TeS3 & 1.300 & 32.399 & mp-1223443 & KBiS2 & 1.416 & 32.126 \\ 
        mp-6215 & AgHgAsS3 & 1.245 & 32.115 & mp-20076 & KCrF6 & 1.398 & 32.128 \\ 
        mp-560067 & AgHgSBr & 1.260 & 32.103 & mp-4026 & KCrS2 & 1.164 & 32.598 \\ 
        mp-23140 & AgHgSI & 1.358 & 32.923 & mp-9263 & KErTe2 & 1.330 & 32.720 \\ 
        mp-580941 & AgI & 1.376 & 32.797 & mp-23582 & KLi6BiO6 & 1.422 & 32.053 \\ 
        mp-3922 & AgSbS2 & 1.372 & 32.894 & mp-7089 & KMgSb & 1.267 & 32.141 \\ 
        mp-2624 & AlSb & 1.226 & 32.400 & mp-6599 & KNb(CuSe2)2 & 1.423 & 32.442 \\ 
        mp-31220 & AlSiTe3 & 1.250 & 32.178 & mp-11740 & KNdTe2 & 1.265 & 32.179 \\ 
        mp-1214793 & AsOsS & 1.180 & 32.580 & mp-5273 & KPrTe2 & 1.280 & 32.180 \\ 
        mp-1228745 & AsRhS & 1.224 & 32.439 & mp-9576 & KSbSe2 & 1.130 & 32.579 \\ 
        mp-505373 & AsSeI & 1.235 & 32.372 & mp-9036 & KSmTe2 & 1.262 & 32.175 \\ 
        mp-7374 & Ba(CuO)2 & 1.386 & 32.837 & mp-1188829 & KTeP2 & 1.209 & 32.474 \\ 
        mp-8278 & Ba(MgP)2 & 1.137 & 32.430 & mp-16763 & KYTe2 & 1.298 & 32.316 \\ 
        mp-1104988 & Ba2Bi2Se3O2 & 1.419 & 32.483 & mp-4841 & LaCuS2 & 1.235 & 32.085 \\ 
        mp-21879 & Ba2CePtO6 & 1.420 & 32.274 & mp-989524 & LaWN3 & 1.210 & 32.111 \\ 
        mp-11902 & Ba2GeSe4 & 1.412 & 32.482 & mp-28989 & Li10BrN3 & 1.367 & 32.778 \\ 
        mp-864638 & Ba2InBiS5 & 1.453 & 32.003 & mp-8181 & Li2CeN2 & 1.204 & 32.518 \\ 
        mp-14448 & Ba2SiTe4 & 1.103 & 32.059 & mp-675779 & Li2Mo3S4 & 1.437 & 32.256 \\ 
        mp-1104423 & Ba2YAg5S6 & 1.366 & 32.625 & mp-7608 & Li2PdO2 & 1.197 & 32.360 \\ 
        mp-19913 & Ba3(InP2)2 & 1.109 & 32.295 & mp-22170 & Li4PbO4 & 1.437 & 32.221 \\ 
        mp-510268 & Cs2Pd3S4 & 1.382 & 32.905 & mp-29365 & Li5BiO5 & 1.360 & 32.142 \\ 
        mp-14338 & Cs2Pt3Se4 & 1.333 & 32.943 & mp-765559 & LiAgF4 & 1.128 & 32.226 \\ 
        mp-505825 & Cs2PtC2 & 1.270 & 32.146 & mp-996959 & LiAuO2 & 1.306 & 32.300 \\ 
        mp-540957 & Cs2TeI6 & 1.388 & 32.872 & mp-31468 & LiCaN & 1.378 & 32.834 \\ 
        mp-10489 & Cs2Ti(CuSe2)2 & 1.364 & 32.423 & mp-1211088 & LiCaP & 1.399 & 32.516 \\ 
        mp-3247 & Cs2TiS3 & 1.272 & 32.111 & mp-1207082 & LiMgSb & 1.444 & 32.123 \\ 
        mp-1112112 & Cs2TlAsBr6 & 1.305 & 32.458 & mp-1029903 & LiNbN2 & 1.142 & 32.257 \\ 
        mp-8684 & Cs2VAgS4 & 1.387 & 32.696 & mp-850189 & LiVSnO4 & 1.452 & 32.070 \\ 
        mp-9856 & Cs2ZrSe3 & 1.369 & 32.197 & mp-10182 & LiZnP & 1.339 & 32.681 \\ 
        mp-505212 & Cs3AuO & 1.326 & 32.832 & mp-1029378 & Mg2SbN3 & 1.358 & 32.718 \\ 
        mp-1096933 & CsAgO2 & 1.106 & 32.028 & mp-5942 & NaAsS2 & 1.355 & 32.182 \\ 
        mp-567913 & CsAuSe3 & 1.192 & 32.556 & mp-23110 & NaHg2IO2 & 1.388 & 32.335 \\ 
        mp-553303 & CsCu3O2 & 1.457 & 32.008 & mp-3744 & NaNbO2 & 1.378 & 32.631 \\ 
        mp-510462 & CsDyZnTe3 & 1.371 & 32.426 & mp-559446 & NaPPdS4 & 1.161 & 32.608 \\ 
        mp-569107 & CsErZnTe3 & 1.386 & 32.198 & mp-8830 & NaRhO2 & 1.377 & 32.889 \\ 
        mp-579556 & CsHoZnTe3 & 1.377 & 32.520 & mp-999447 & NaSmTe2 & 1.174 & 32.432 \\ 
        mp-11124 & CsLaHgSe3 & 1.455 & 32.029 & mp-13275 & NaSrP & 1.236 & 32.247 \\ 
        mp-34034 & Ba8P5Br & 1.164 & 32.624 & mp-5475 & NaTaN2 & 1.276 & 32.105 \\ 
        mp-9195 & BaCuSeF & 1.427 & 32.163 & mp-1180119 & NaTiCu3Se4 & 1.371 & 32.894 \\ 
        mp-28007 & BaHgS2 & 1.157 & 32.229 & mp-4043 & NbCu3Se4 & 1.360 & 32.877 \\ 
        mp-1214348 & BaLaAgTe3 & 1.293 & 32.274 & mp-1016197 & MgSiAs2 & 1.303 & 32.007 \\ 
        mp-27251 & Bi2TeO6 & 1.259 & 32.216 & mp-2604 & MgTe2 & 1.116 & 32.453 \\ 
        mp-33723 & BiTeBr & 1.236 & 32.304 & mp-19442 & Mn(Ni3O4)2 & 1.425 & 32.400 \\ 
        mp-22965 & BiTeI & 1.233 & 32.383 & mp-19142 & Mn2V2O7 & 1.204 & 32.210 \\ 
        mp-1078908 & Ca2CdP2 & 1.293 & 32.279 & mp-28013 & MnI2 & 1.171 & 32.439 \\ 
        mp-1205338 & Ca3(AlAs2)2 & 1.110 & 32.339 & mp-1634 & MoSe2 & 1.301 & 32.428 \\ 
        mp-1205337 & Ca3(AlP2)2 & 1.248 & 32.276 & mp-3622 & Na2PdF4 & 1.139 & 32.453 \\ 
        mp-8789 & Ca4As2O & 1.213 & 32.457 & mp-3527 & Na3AgO2 & 1.144 & 32.343 \\ 
        mp-5380 & Ca4P2O & 1.271 & 32.169 & mp-5122 & Na3AlP2 & 1.346 & 32.727 \\ 
        mp-28879 & Ca5P8 & 1.142 & 32.352 & mp-28400 & P3Ru & 1.145 & 32.580 \\ 
        mp-1213950 & CaAgAsO4 & 1.357 & 32.508 & mp-1103842 & P4Os & 1.304 & 32.446 \\ 
        mp-1029633 & CaSnN2 & 1.377 & 32.899 & mp-27173 & P4Ru & 1.259 & 32.221 \\ 
        mp-4666 & CdSiP2 & 1.426 & 32.353 & mp-1219924 & PRhS & 1.399 & 32.737 \\ 
        mp-19178 & CoAgO2 & 1.332 & 32.949 & mp-9798 & Rb(SbSe2)2 & 1.119 & 32.499 \\ 
        mp-1226003 & CoPS & 1.123 & 32.533 & mp-1111655 & Rb2AgAsCl6 & 1.295 & 32.155 \\ 
        mp-551407 & CoSeO4 & 1.366 & 32.528 & mp-1110602 & Rb2AgRhF6 & 1.390 & 32.612 \\ 
        mp-21074 & Cr2HgO4 & 1.258 & 32.223 & mp-1205560 & Rb2As2Pd & 1.146 & 32.623 \\ 
        mp-21355 & Cr2TeO6 & 1.424 & 32.435 & mp-1206059 & Rb2NaCrCl6 & 1.363 & 32.534 \\ 
        mp-996996 & CrAuO2 & 1.266 & 32.199 & mp-1205714 & Rb2P2Pd & 1.265 & 32.199 \\ 
        mp-8890 & Cs(SbS2)2 & 1.441 & 32.209 & mp-11695 & Rb2Pd3S4 & 1.185 & 32.591 \\ 
        mp-3312 & Cs(SbSe2)2 & 1.122 & 32.519 & mp-28145 & Rb2PdCl6 & 1.287 & 32.203 \\ 
        mp-1113578 & Cs2AgAsBr6 & 1.116 & 32.412 & mp-10919 & Rb2PtC2 & 1.136 & 32.607 \\ 
        mp-1106156 & Cs2Hg3Se4 & 1.195 & 32.199 & mp-8901 & Rb2VAgS4 & 1.332 & 32.777 \\ 
        mp-1096926 & Cs2InAgCl6 & 1.342 & 32.381 & mp-15219 & Rb2VCuS4 & 1.244 & 32.164 \\ 
        mp-510065 & CsNdHgSe3 & 1.354 & 32.775 & mp-9718 & Rb3BAs2 & 1.294 & 32.236 \\ 
        mp-11742 & CsNdTe2 & 1.292 & 32.246 & mp-9274 & Rb3Sb2Au3 & 1.222 & 32.443 \\ 
        mp-12342 & CsNdZnTe3 & 1.365 & 32.844 & mp-8603 & RbAgO & 1.417 & 32.273 \\ 
        mp-7211 & CsPrHgSe3 & 1.337 & 32.836 & mp-30041 & RbBiS2 & 1.339 & 32.409 \\ 
        mp-12341 & CsPrZnTe3 & 1.361 & 32.811 & mp-9845 & RbCaAs & 1.299 & 32.385 \\ 
        mp-2969 & CsSbSe2 & 1.131 & 32.582 & mp-5808 & RbNdTe2 & 1.299 & 32.364 \\ 
        mp-7212 & CsSmHgSe3 & 1.363 & 32.852 & mp-616564 & RbPbIO6 & 1.375 & 32.893 \\ 
        mp-12343 & CsSmZnTe3 & 1.409 & 32.570 & mp-999269 & RbSmTe2 & 1.316 & 32.610 \\ 
        mp-1542038 & CsSnSe3 & 1.190 & 32.344 & mp-9008 & RbTeAu & 1.108 & 32.155 \\ 
        mp-638078 & CsTbZnTe3 & 1.357 & 32.619 & mp-16764 & RbYTe2 & 1.356 & 32.816 \\ 
        mp-1541909 & CsTeAu & 1.416 & 32.543 & mp-28918 & Sb2WO6 & 1.189 & 32.103 \\ 
        mp-1205877 & CsTlO & 1.391 & 32.556 & mp-11178 & SrCeN2 & 1.268 & 32.198 \\ 
        mp-1103187 & CsTmZnTe3 & 1.384 & 32.242 & mp-13276 & SrLiP & 1.343 & 32.859 \\ 
        mp-11123 & CsYHgSe3 & 1.395 & 32.705 & mp-1542758 & SrMoO3 & 1.309 & 32.156 \\ 
        mp-1103744 & CsYZnTe3 & 1.405 & 32.475 & mp-1029275 & SrZrN2 & 1.227 & 32.414 \\ 
        mp-1025340 & Cu2WSe4 & 1.276 & 32.199 & mp-9295 & TaCu3Te4 & 1.141 & 32.597 \\ 
        mp-1225724 & Dy(CuS)3 & 1.397 & 32.764 & mp-1217848 & Tb(CuS)3 & 1.341 & 32.942 \\ 
        mp-1095296 & Dy4Te3S4 & 1.397 & 32.740 & mp-30291 & Tb4Se3N2 & 1.345 & 32.318 \\ 
        mp-1181462 & DyAgSe2 & 1.306 & 32.049 & mp-5737 & TbCuS2 & 1.183 & 32.528 \\ 
        mp-1225120 & Er(CuS)3 & 1.446 & 32.155 & mp-9481 & TcS2 & 1.189 & 32.435 \\ 
        mp-1189719 & Er3Tl2Cu5S8 & 1.283 & 32.197 & mp-27741 & TeAuI & 1.121 & 32.518 \\ 
        mp-22421 & Fe2GeO4 & 1.408 & 32.472 & mp-1217271 & Th2PNO & 1.313 & 32.545 \\ 
        mp-35596 & Fe2NiO4 & 1.362 & 32.772 & mp-1079673 & ThSe3 & 1.273 & 32.019 \\ 
        mp-19225 & FeAgO2 & 1.148 & 32.417 & mp-29091 & Ti(CuS)4 & 1.426 & 32.411 \\ 
        mp-30946 & Ga2PdI8 & 1.418 & 32.494 & mp-29337 & Tl3BS3 & 1.422 & 32.188 \\ 
        mp-541785 & GePdS3 & 1.344 & 32.944 & mp-28490 & Tl3BSe3 & 1.225 & 32.277 \\ 
        mp-2242 & GeS & 1.238 & 32.358 & mp-8630 & SbIrS & 1.396 & 32.772 \\ 
        mp-29419 & Hf(Te2Cl3)2 & 1.263 & 32.025 & mp-1095507 & SbIrSe & 1.147 & 32.626 \\ 
        mp-985829 & HfS2 & 1.224 & 32.260 & mp-1102833 & SbOsS & 1.119 & 32.469 \\ 
        mp-554921 & Hg(BiS2)2 & 1.186 & 32.113 & mp-1101771 & SbOsSe & 1.123 & 32.518 \\ 
        mp-1224171 & Hg11I2BrClO4 & 1.308 & 32.508 & mp-1209072 & SbSeBr & 1.438 & 32.268 \\ 
        mp-28875 & Hg2P3Cl & 1.135 & 32.326 & mp-22996 & SbSeI & 1.381 & 32.905 \\ 
        mp-23192 & HgI2 & 1.332 & 32.915 & mp-1094066 & ScAgS2 & 1.238 & 32.114 \\ 
        mp-9006 & Ho2CF2 & 1.136 & 32.584 & mp-1206699 & Si4P4Os & 1.434 & 32.318 \\ 
        mp-27988 & IF7 & 1.440 & 32.184 & mp-1863 & SiAs & 1.452 & 32.082 \\ 
        mp-22186 & In6Ge2PtO9 & 1.345 & 32.905 & mp-5081 & SmCuS2 & 1.168 & 32.548 \\ 
        mp-1223929 & InCuGeS4 & 1.135 & 32.121 & mp-550820 & SmZnPO & 1.361 & 32.279 \\ 
        mp-2437 & IrF3 & 1.141 & 32.616 & mp-36381 & Sn(PS3)2 & 1.194 & 32.053 \\ 
        mp-9797 & K(SbSe2)2 & 1.117 & 32.474 & mp-37091 & Sr(AlTe2)2 & 1.445 & 32.156 \\ 
        mp-28769 & K(SnSe2)2 & 1.173 & 32.605 & mp-863260 & Sr(MgAs)2 & 1.346 & 32.462 \\ 
        mp-9778 & K2AgP & 1.232 & 32.377 & mp-1986938 & Sr2Bi2Se3O2 & 1.387 & 32.825 \\ 
        mp-7643 & K2AgSb & 1.187 & 32.586 & mp-1208839 & Sr2H6Pt & 1.276 & 32.204 \\ 
        mp-1110871 & K2CuBiCl6 & 1.162 & 32.627 & mp-1189305 & Sr3(AlAs2)2 & 1.146 & 32.614 \\ 
        mp-8446 & K2CuP & 1.142 & 32.414 & mp-9843 & Sr3(AlP2)2 & 1.337 & 32.972 \\ 
        mp-1206266 & K2NaCrCl6 & 1.388 & 32.263 & mp-8299 & Sr4As2O & 1.192 & 32.530 \\ 
        mp-23067 & K2PdCl6 & 1.203 & 32.191 & mp-8298 & Sr4P2O & 1.213 & 32.427 \\ 
        mp-1068941 & K2PdS2 & 1.117 & 32.229 & mp-676540 & TlSbS2 & 1.267 & 32.171 \\ 
        mp-1062676 & K2Pt & 1.315 & 32.170 & mp-1821 & WSe2 & 1.447 & 32.157 \\ 
        mp-8235 & K2SiP2 & 1.209 & 32.262 & mp-1216184 & Y(CuS)3 & 1.420 & 32.502 \\ 
        mp-8965 & K2Sn2S5 & 1.406 & 32.509 & mp-1188559 & YAgSe2 & 1.280 & 32.144 \\ 
        mp-8900 & K2VAgS4 & 1.305 & 32.285 & mp-35311 & ZnCrF6 & 1.246 & 32.298 \\ 
        mp-15147 & K2VCuS4 & 1.234 & 32.213 & mp-1215793 & ZnCrFeO4 & 1.323 & 32.260 \\ 
        mp-15220 & K2VCuSe4 & 1.125 & 32.531 & mp-4524 & ZnGeP2 & 1.178 & 32.533 \\ 
        mp-14206 & K3Ag3As2 & 1.253 & 32.257 & mp-13983 & ZnPdF6 & 1.357 & 32.892 \\ 
        mp-28347 & K3Al2As3 & 1.186 & 32.425 & mp-541912 & ZrBrN & 1.442 & 32.206 \\ 
        mp-14205 & K3Cu3As2 & 1.274 & 32.206 & mp-1100415 & ZrSbRh & 1.165 & 32.628 \\ 
\end{longtblr}
}

\subsection{High-quantum-weight quantum material}

{\small
\begin{longtblr}
[
caption = {The list of 296 high-quantum-weight $K_{xx}$ materials with the Materials Project IDs (MP-ID). $K_{xx}$ in unit of $h/e^2$ and the energy band gap $E_g$ in the unit of eV.},
label = {tab:2}
]
{
colspec={|l|l|l|l||l|l|l|l|},
hline{1,Z}={1}{-}{0.05em},
hline{1,Z}={2}{-}{0.05em},
hline{2}={0.05em}
}
MP-ID & Formula & $E_g$ & $K_{xx}$ & MP-ID & Formula & $E_g$ & $K_{xx}$ \\
        mp-10910 & Al2Ru & 0.086 & 28.617 & mp-1219954 & NiSb6Ru & 0.068 & 32.614 \\ 
        mp-1228817 & AlReGe & 0.119 & 35.783 & mp-1209793 & Np(FeP3)4 & 0.480 & 33.162 \\ 
        mp-1228809 & AlReSi & 0.047 & 29.443 & mp-10155 & P2Ir & 0.633 & 46.937 \\ 
        mp-15649 & As2Ir & 0.816 & 44.405 & mp-2319 & P2Os & 0.890 & 40.052 \\ 
        mp-2455 & As2Os & 0.665 & 39.786 & mp-28266 & P2Pd & 0.342 & 32.287 \\ 
        mp-2513 & As2Pt & 0.079 & 34.342 & mp-730 & P2Pt & 1.019 & 33.680 \\ 
        mp-15954 & As2Rh & 0.263 & 38.237 & mp-15953 & P2Rh & 0.370 & 39.380 \\ 
        mp-766 & As2Ru & 0.449 & 39.648 & mp-1413 & P2Ru & 0.479 & 38.024 \\ 
        mp-540912 & As3Ir & 0.027 & 40.000 & mp-13853 & P3Ir & 0.085 & 38.772 \\ 
        mp-1228823 & AsIrS & 1.863 & 42.819 & mp-28400 & P3Ru & 1.145 & 33.822 \\ 
        mp-1228810 & AsIrSe & 1.470 & 44.948 & mp-1103842 & P4Os & 1.304 & 35.341 \\ 
        mp-1214793 & AsOsS & 1.180 & 38.644 & mp-27173 & P4Ru & 1.259 & 33.051 \\ 
        mp-1214784 & AsOsSe & 1.003 & 40.793 & mp-1220004 & PIrS & 1.865 & 39.670 \\ 
        mp-1228745 & AsRhS & 1.224 & 36.956 & mp-1102534 & POsS & 0.944 & 32.364 \\ 
        mp-1228724 & AsRhSe & 0.914 & 39.136 & mp-1219924 & PRhS & 1.399 & 33.787 \\ 
        mp-1214786 & AsRuS & 0.915 & 37.758 & mp-1102531 & PRhSe & 1.083 & 38.673 \\ 
        mp-160 & B & 1.433 & 28.929 & mp-2201 & PbSe & 0.426 & 28.452 \\ 
        mp-1227860 & BaPPd & 0.296 & 35.169 & mp-1206667 & PrNiBi & 0.334 & 30.592 \\ 
        mp-1009084 & BeSnAs2 & 0.694 & 28.774 & mp-999305 & PrRh & 0.176 & 31.660 \\ 
        mp-675543 & Bi2PbSe4 & 0.455 & 28.749 & mp-1209288 & PrTeAs & 0.124 & 28.508 \\ 
        mp-27910 & Bi2Te2S & 0.437 & 38.461 & mp-288 & PtS & 0.385 & 29.370 \\ 
        mp-29666 & Bi2Te2Se & 0.544 & 36.650 & mp-1115 & PtSe2 & 0.619 & 28.405 \\ 
        mp-1102836 & BiIrS & 0.689 & 45.857 & mp-1219672 & Rb(Nb3Se4)2 & 0.399 & 30.326 \\ 
        mp-1103228 & BiIrSe & 0.535 & 46.385 & mp-17401 & Rb3Sn4Au & 0.581 & 36.759 \\ 
        mp-1103098 & BiRhS & 0.340 & 37.728 & mp-5222 & ReTeS & 0.075 & 31.941 \\ 
        mp-1101765 & BiRhSe & 0.238 & 38.730 & mp-1922 & RuSe2 & 0.314 & 32.891 \\ 
        mp-1095302 & BiSe2 & 0.672 & 27.910 & mp-1247 & Sb2Ir & 0.510 & 41.249 \\ 
        mp-1227426 & BiTeRh & 0.162 & 36.360 & mp-2695 & Sb2Os & 0.345 & 36.534 \\ 
        mp-11918 & Ca(BeN)2 & 2.077 & 28.484 & mp-20928 & Sb2Ru & 0.003 & 37.206 \\ 
        mp-11168 & Ca(PIr)2 & 0.494 & 36.158 & mp-3525 & Sb2Te2Se & 0.131 & 53.130 \\ 
        mp-866229 & Ca2SnHg & 0.149 & 29.260 & mp-1201 & Sb2Te3 & 0.131 & 37.118 \\ 
        mp-31149 & Ca3BiN & 0.369 & 30.271 & mp-8612 & Sb2TeSe2 & 0.486 & 42.499 \\ 
        mp-989590 & Ca6Sn2NF & 0.132 & 30.401 & mp-867249 & TaAlFe2 & 0.324 & 37.970 \\ 
        mp-756301 & Cd(CoO2)2 & 0.293 & 30.160 & mp-12561 & TaAs2 & 0.013 & 37.663 \\ 
        mp-1021508 & Ce(As3Ru)4 & 0.238 & 43.599 & mp-867507 & TaGaFe2 & 0.060 & 39.826 \\ 
        mp-1188930 & Ce(BOs)4 & 0.227 & 35.037 & mp-31454 & TaSbRu & 0.665 & 37.112 \\ 
        mp-1021509 & Ce(FeAs3)4 & 0.147 & 46.688 & mp-1100408 & TaSnRh & 1.059 & 36.916 \\ 
        mp-16272 & Ce(FeP3)4 & 0.521 & 41.533 & mp-28691 & TaTe4I & 0.563 & 31.268 \\ 
        mp-1181682 & Ce(FeSb3)4 & 0.231 & 41.049 & mp-1206686 & TbNiBi & 0.230 & 29.679 \\ 
        mp-38564 & Ce(Mo3S4)2 & 0.338 & 38.439 & mp-3716 & TbNiSb & 0.331 & 29.431 \\ 
        mp-1021505 & Ce(P3Os)4 & 0.283 & 39.770 & mp-16313 & TbSbPt & 0.097 & 28.091 \\ 
        mp-10069 & Ce(P3Ru)4 & 0.192 & 38.990 & mp-28029 & TcP3 & 0.451 & 34.375 \\ 
        mp-1105726 & Ce(Sb3Os)4 & 0.051 & 41.596 & mp-19 & Te & 0.186 & 36.109 \\ 
        mp-1189811 & Ce(Sb3Ru)4 & 0.043 & 38.882 & mp-602 & Te2Mo & 0.863 & 60.792 \\ 
        mp-1069558 & Ce2SeN2 & 0.948 & 28.008 & mp-267 & Te2Ru & 0.285 & 47.772 \\ 
        mp-10478 & CeAsRh & 0.026 & 39.711 & mp-484 & Te3As2 & 0.433 & 32.130 \\ 
        mp-1213954 & CeBiRh & 0.014 & 36.873 & mp-1217418 & Te4Mo2Ru & 0.250 & 42.626 \\ 
        mp-1226908 & CeCuGeH & 0.201 & 38.067 & mp-29190 & Te4MoBr & 0.858 & 43.505 \\ 
        mp-1076951 & CeNiGe & 0.057 & 39.243 & mp-1217272 & TeAsIr & 1.085 & 49.334 \\ 
        mp-2109 & CePt2 & 0.009 & 47.908 & mp-1095634 & TeAsRu & 0.825 & 45.580 \\ 
        mp-19178 & CoAgO2 & 1.332 & 31.229 & mp-19717 & TePb & 0.806 & 31.044 \\ 
        mp-2715 & CoAs2 & 0.172 & 39.226 & mp-22945 & TeRhCl & 0.780 & 34.465 \\ 
        mp-14285 & CoP2 & 0.438 & 37.981 & mp-1217322 & TeSe & 0.706 & 28.630 \\ 
        mp-1226003 & CoPS & 1.123 & 31.190 & mp-1189150 & Th(As3Os)4 & 0.466 & 41.001 \\ 
        mp-1317 & CoSb3 & 0.163 & 31.824 & mp-1179056 & Th(BOs)4 & 0.243 & 32.308 \\ 
        mp-1226296 & CrCu(PSe3)2 & 0.640 & 27.964 & mp-9619 & Th(FeP3)4 & 0.535 & 38.004 \\ 
        mp-23116 & CuBiSeO & 0.301 & 31.968 & mp-1208278 & Th(P3Ru)4 & 0.075 & 36.143 \\ 
        mp-927 & CuP2 & 0.868 & 30.106 & mp-382 & Th3As4 & 0.271 & 33.414 \\ 
        mp-20331 & CuSbSe2 & 0.499 & 29.010 & mp-23270 & Th3Bi4 & 0.092 & 31.759 \\ 
        mp-30452 & DyNiBi & 0.208 & 29.914 & mp-1347 & Th3P4 & 0.262 & 29.339 \\ 
        mp-4510 & DyNiSb & 0.312 & 29.328 & mp-552 & Th3Sb4 & 0.038 & 35.202 \\ 
        mp-1225491 & DySbPd & 0.230 & 28.246 & mp-22786 & ThNiSn & 0.268 & 38.779 \\ 
        mp-16327 & DySbPt & 0.049 & 28.041 & mp-10623 & ThSbRh & 0.744 & 34.373 \\ 
        mp-1212803 & DyTeAs & 0.329 & 30.294 & mp-19886 & ThSnPt & 0.674 & 36.899 \\ 
        mp-1206712 & ErNiBi & 0.170 & 29.991 & mp-3718 & ThTeO & 0.242 & 31.685 \\ 
        mp-21272 & ErNiSb & 0.274 & 29.580 & mp-1217120 & Ti2FeNiSb2 & 0.941 & 35.129 \\ 
        mp-11836 & ErSbPd & 0.209 & 28.404 & mp-998980 & TiAlFeCo & 0.125 & 36.827 \\ 
        mp-5640 & ErSnAu & 0.006 & 29.391 & mp-5967 & TiCoSb & 1.043 & 34.820 \\ 
        mp-1212634 & ErTeAs & 0.362 & 30.825 & mp-866375 & TiFe2Ge & 0.171 & 47.547 \\ 
        mp-21386 & Eu(PIr)2 & 0.336 & 29.941 & mp-866141 & TiFe2Si & 0.402 & 43.603 \\ 
        mp-1095411 & EuIn2(GeIr)4 & 0.127 & 41.390 & mp-19963 & TiFe2Sn & 0.049 & 46.711 \\ 
        mp-9198 & Fe(SiP)4 & 1.023 & 32.851 & mp-961673 & TiFeTe & 0.985 & 35.642 \\ 
        mp-35596 & Fe2NiO4 & 1.362 & 28.153 & mp-1008680 & TiGePt & 0.882 & 34.051 \\ 
        mp-19225 & FeAgO2 & 1.148 & 28.932 & mp-574169 & TiGeTe6 & 0.337 & 36.452 \\ 
        mp-2008 & FeAs2 & 0.283 & 43.455 & mp-924130 & TiNiSn & 0.453 & 31.398 \\ 
        mp-561511 & FeAsS & 0.737 & 44.591 & mp-20459 & TiPbO3 & 1.813 & 29.328 \\ 
        mp-1101894 & FeAsSe & 0.449 & 47.376 & mp-961682 & TiSnPd & 0.482 & 34.096 \\ 
        mp-1101938 & 2-Feb & 0.515 & 40.344 & mp-30847 & TiSnPt & 0.868 & 32.578 \\ 
        mp-1224908 & FeNiSb6 & 0.128 & 33.806 & mp-29711 & Tl2Te3 & 0.639 & 28.265 \\ 
        mp-1274279 & FeO & 1.816 & 30.190 & mp-1239 & Sb3Ir & 0.054 & 37.580 \\ 
        mp-20027 & FeP2 & 0.433 & 45.520 & mp-1219478 & SbAsRh & 0.539 & 37.103 \\ 
        mp-1101971 & FePS & 0.505 & 36.542 & mp-8630 & SbIrS & 1.396 & 47.964 \\ 
        mp-1522 & FeS2 & 0.877 & 28.015 & mp-1095507 & SbIrSe & 1.147 & 50.175 \\ 
        mp-1224896 & FeSb6Pd & 0.249 & 38.417 & mp-1102833 & SbOsS & 1.119 & 35.894 \\ 
        mp-27904 & FeSbS & 0.547 & 44.702 & mp-1101771 & SbOsSe & 1.123 & 37.484 \\ 
        mp-1103256 & FeSbSe & 0.622 & 47.452 & mp-1103317 & SbRhS & 0.872 & 40.061 \\ 
        mp-1102580 & FeSbTe & 0.298 & 45.590 & mp-1102366 & SbRhSe & 0.403 & 41.668 \\ 
        mp-760 & FeSe2 & 0.374 & 32.860 & mp-1103434 & SbRuS & 0.925 & 35.885 \\ 
        mp-871 & FeSi & 0.180 & 48.738 & mp-1102395 & SbRuSe & 0.783 & 37.385 \\ 
        mp-19880 & FeTe2 & 0.077 & 53.925 & mp-1102430 & SbTeIr & 0.695 & 48.479 \\ 
        mp-1102288 & FeTeAs & 0.682 & 49.792 & mp-1103213 & SbTeOs & 0.858 & 40.618 \\ 
        mp-1072429 & Ga2Ru & 0.127 & 37.377 & mp-1219458 & SbTeRh & 0.552 & 43.280 \\ 
        mp-570844 & Ga3Os & 0.677 & 32.340 & mp-1102857 & SbTeRu & 0.627 & 41.492 \\ 
        mp-14791 & Ge2Te5As2 & 0.409 & 36.101 & mp-15661 & Sc4C3 & 0.329 & 29.023 \\ 
        mp-541312 & Ge3(Te3As)2 & 0.343 & 36.303 & mp-1219274 & ScNb(NiSn)2 & 0.425 & 39.440 \\ 
        mp-9548 & GeAs & 0.511 & 32.173 & mp-30459 & ScNiBi & 0.187 & 30.051 \\ 
        mp-1095275 & GeP & 0.489 & 32.531 & mp-3432 & ScNiSb & 0.268 & 33.048 \\ 
        mp-1101755 & GePtS & 1.020 & 34.701 & mp-569779 & ScSbPd & 0.016 & 32.624 \\ 
        mp-20817 & GePtSe & 0.660 & 36.222 & mp-7173 & ScSbPt & 0.644 & 31.513 \\ 
        mp-938 & GeTe & 0.809 & 38.465 & mp-2894 & ScSnAu & 0.122 & 31.638 \\ 
        mp-1224373 & GeTePt & 0.256 & 42.885 & mp-1100405 & ScTeRh & 0.385 & 32.502 \\ 
        mp-1029330 & Hf2SeN2 & 0.607 & 29.039 & mp-988 & Si3N4 & 4.250 & 29.466 \\ 
        mp-567817 & HfGeTe4 & 0.331 & 35.332 & mp-29157 & Si3P2Pt & 0.607 & 31.683 \\ 
        mp-924128 & HfNiSn & 0.387 & 33.746 & mp-1206699 & Si4P4Os & 1.434 & 31.817 \\ 
        mp-1224358 & HfSb4Mo & 0.139 & 40.430 & mp-14983 & Si4P4Ru & 1.460 & 31.317 \\ 
        mp-866062 & HfSiRu2 & 0.169 & 41.411 & mp-2488 & SiOs & 0.512 & 46.524 \\ 
        mp-11869 & HfSnPd & 0.034 & 34.777 & mp-1103261 & SiPtSe & 0.788 & 32.805 \\ 
        mp-1212255 & Ho4Ga12Pt & 0.219 & 27.898 & mp-189 & SiRu & 0.230 & 46.453 \\ 
        mp-1018139 & HoNiBi & 0.194 & 30.141 & mp-13305 & SmSnAu & 0.066 & 29.020 \\ 
        mp-4174 & HoNiSb & 0.292 & 29.461 & mp-1208842 & SmTeAs & 0.380 & 29.450 \\ 
        mp-1212072 & HoPPd & 0.018 & 30.273 & mp-38605 & Sn(BiTe2)2 & 0.427 & 30.676 \\ 
        mp-30390 & HoSnAu & 0.022 & 29.527 & mp-27947 & Sn(SbTe2)2 & 0.253 & 37.274 \\ 
        mp-1077901 & InCo3SnS2 & 0.197 & 36.997 & mp-1218953 & SnGe4Te4Se & 0.613 & 37.954 \\ 
        mp-1223754 & K(Nb3Se4)2 & 0.391 & 30.122 & mp-1218931 & SnPtS & 0.818 & 36.721 \\ 
        mp-18500 & K3Sn4Au & 0.397 & 38.674 & mp-1218926 & SnPtSe & 0.574 & 38.511 \\ 
        mp-11132 & KPrTe4 & 0.176 & 28.023 & mp-691 & SnSe & 0.516 & 29.985 \\ 
        mp-1206717 & LaBiPd & 0.122 & 29.549 & mp-1883 & SnTe & 0.042 & 40.639 \\ 
        mp-10288 & LaCuTeS & 0.577 & 28.022 & mp-1218898 & SnTePt & 0.090 & 38.769 \\ 
        mp-1002107 & LaRh & 0.081 & 29.213 & mp-15074 & Sr(PIr)2 & 0.442 & 43.977 \\ 
        mp-550514 & LaTaN2O & 0.671 & 29.423 & mp-567278 & Sr(Si3N4)2 & 3.243 & 27.998 \\ 
        mp-989524 & LaWN3 & 1.210 & 31.927 & mp-9379 & Sr(SnAs)2 & 0.009 & 29.791 \\ 
        mp-1222744 & LaZnCuP2 & 0.186 & 28.513 & mp-29662 & TlBiSe2 & 0.232 & 27.943 \\ 
        mp-675779 & Li2Mo3S4 & 1.437 & 29.217 & mp-27438 & TlBiTe2 & 0.468 & 29.432 \\ 
        mp-1029385 & Li2SnN2 & 1.697 & 34.949 & mp-4573 & TlSbTe2 & 0.127 & 33.736 \\ 
        mp-1185329 & LiAcRh2 & 0.296 & 34.203 & mp-568269 & TmNiBi & 0.149 & 29.278 \\ 
        mp-569450 & LiB6C & 1.392 & 28.152 & mp-4025 & TmNiSb & 0.255 & 29.330 \\ 
        mp-7936 & LiNbS2 & 0.713 & 27.955 & mp-1776 & UN2 & 0.693 & 31.443 \\ 
        mp-1025496 & LiNbSe2 & 0.732 & 28.673 & mp-567636 & VFeSb & 0.349 & 41.101 \\ 
        mp-30457 & LuNiBi & 0.013 & 30.100 & mp-30460 & YNiBi & 0.202 & 30.168 \\ 
        mp-20185 & LuNiSb & 0.215 & 31.174 & mp-11520 & YNiSb & 0.295 & 29.581 \\ 
        mp-11917 & Mg(BeN)2 & 4.066 & 32.452 & mp-1207056 & YSbPd & 0.474 & 28.380 \\ 
        mp-865280 & NbAlFe2 & 0.321 & 40.546 & mp-4964 & YSbPt & 0.110 & 28.334 \\ 
        mp-1094088 & NbCoSn & 0.970 & 37.556 & mp-4697 & Zn(CrSe2)2 & 0.039 & 29.066 \\ 
        mp-9437 & NbFeSb & 0.514 & 40.147 & mp-1215609 & ZnGa11Co4 & 0.240 & 32.307 \\ 
        mp-977410 & NbGaFe2 & 0.091 & 41.533 & mp-753 & ZnSb & 0.028 & 32.248 \\ 
        mp-9339 & NbP & 0.116 & 35.584 & mp-1020712 & ZnSiN2 & 3.194 & 30.130 \\ 
        mp-1969 & NbSb2 & 0.005 & 35.669 & mp-11583 & Zr2SN2 & 0.564 & 28.566 \\ 
        mp-505297 & NbSbRu & 0.377 & 38.616 & mp-1079726 & Zr2SeN2 & 0.337 & 30.393 \\ 
        mp-864954 & MgMoN2 & 0.741 & 31.584 & mp-1093991 & ZrAsIr & 0.269 & 37.090 \\ 
        mp-864908 & MgTiIr2 & 0.013 & 42.230 & mp-31451 & ZrCoBi & 0.977 & 35.696 \\ 
        mp-1104183 & Mn(BiSe2)2 & 0.335 & 28.012 & mp-1095610 & ZrGePt & 0.088 & 40.359 \\ 
        mp-18750 & Mn(FeO2)2 & 0.998 & 28.400 & mp-13542 & ZrGeTe4 & 0.376 & 37.612 \\ 
        mp-569859 & MnTl2SnTe4 & 0.310 & 28.657 & mp-924129 & ZrNiSn & 0.495 & 35.101 \\ 
        mp-1221499 & Mo2RuSe4 & 0.632 & 28.220 & mp-1100415 & ZrSbRh & 1.165 & 36.074 \\ 
        mp-1634 & MoSe2 & 1.301 & 30.583 & mp-1183042 & ZrSiRu2 & 0.235 & 42.829 \\ 
        mp-1008858 & NdBiPd & 0.105 & 30.821 & mp-961687 & ZrSnPd & 0.482 & 35.931 \\ 
        mp-1206278 & NdInCu4 & 0.005 & 33.139 & mp-961713 & ZrSnPt & 0.963 & 34.645 \\ 
        mp-1206719 & NdNiBi & 0.512 & 30.028 & mp-605 & ZrTe5 & 0.019 & 33.901 \\ 
\end{longtblr}
}